\documentclass[fleqn,usenatbib]{mnras}

\usepackage{newtxtext, newtxmath}

\usepackage[T1]{fontenc}

\DeclareRobustCommand{\VAN}[3]{#2}
\let\VANthebibliography\thebibliography
\def\thebibliography{\DeclareRobustCommand{\VAN}[3]{##3}\VANthebibliography}

\usepackage{graphicx}	
\usepackage{amsmath}	

\title[M\&L Generalisation to SR Transients]{Generalisation of the Menegozzi \& Lamb Maser Algorithm to the Transient Superradiance Regime}

\author[C. M. Wyenberg et al.]{C. M. Wyenberg,$^{1}$\thanks{E-mail: cwyenber@uwo.ca} B. Lankhaar,$^{2}$ F. Rajabi,$^{3,4}$ M. A. Chamma,$^{1}$ and M. Houde$^{1}$\thanks{E-mail: mhoude2@uwo.ca}\\
$^{1}$Department of Physics and Astronomy, The University of Western Ontario, 1151 Richmond Street, London, Ontario N6A 3K7, Canada\\
$^{2}$Department of Space, Earth and Environment, Chalmers University of Technology, Onsala Space Observatory, 439 92 Onsala, Sweden\\
$^{3}$Perimeter Institute for Theoretical Physics, Waterloo, ON N2L 2Y5, Canada\\
$^{4}$Institute for Quantum Computing and Department of Physics and Astronomy, The University of Waterloo, 200 University Ave. West, \\Waterloo, Ontario N2L 3G1, Canada\\}

\date{}

\pubyear{2021}

\begin{document}
\label{firstpage}
\pagerange{\pageref{firstpage}--\pageref{lastpage}}
\maketitle

\begin{abstract}
We investigate the application of the conventional quasi-steady state maser modelling algorithm of Menegozzi \& Lamb (ML) to the high field transient regime of the one-dimensional Maxwell-Bloch (MB) equations for a velocity distribution of atoms or molecules. We quantify the performance of a first order perturbation approximation available within the ML framework when modelling regions of increasing electric field strength,  and we show that the ML algorithm is unable to accurately describe the key transient features of R. H. Dicke's superradiance (SR). We extend the existing approximation to one of variable fidelity, and we derive a generalisation of the ML algorithm convergent in the transient SR regime by performing an integration on the MB equations prior to their Fourier representation. We obtain a manifestly unique integral Fourier representation of the MB equations which is $\mathcal{O}\left(N\right)$ complex in the number of velocity channels $N$ and which is capable of simulating transient SR processes at varying degrees of fidelity. As a proof of operation, we demonstrate our algorithm's accuracy against reference time domain simulations of the MB equations for transient SR responses to the sudden inversion of a sample possessing a velocity distribution of moderate width. We investigate the performance of our algorithm at varying degrees of approximation fidelity, and we prescribe fidelity requirements for future work simulating SR processes across wider velocity distributions.
\end{abstract}


\section{Introduction}\label{sec:introduction}

The theory of quantum electrodynamics (QED) describes the emission of a photon from an excited atom or molecule\footnote{For brevity, we shall herein refer to "molecules" only; however, all discussion and results to follow apply equally well to either atoms or molecules.} through its interaction with the quantized radiation field. More generally, a large collection of molecules interacting with their common radiation field can produce complex spontaneous radiative phenomena. In the well-known process of microwave amplification by stimulated emission of radiation, for example, the presence of a photon in the radiation field as generated from one molecule enhances the emission rate from adjacent molecules. In an ideal system, in which stimulated emission occurs between molecules of similar velocities and without dephasing interactions, stimulated photons possess frequencies very near those of the stimulating photons. Such processes lead to a high degree of coherence within the radiation field. Conversely, a realistic amplification by stimulated emission process is usually only weakly coherent, in that broad velocity distributions as well as collisional and other dephasing processes together determine the quasi-steady state radiation field, population inversion level, and polarisation profiles \citep{Rajabi2020}. In astrophysics, microwave amplification by stimulated emission of radiation (maser) processes have been observed in regions containing sufficiently velocity-coherent gases of molecules \citep{Elitzur1992,Gray2012}.

There also exists a cooperative coherent spontaneous emission process related to (but distinct from) amplification by stimulated emission, known as R. H. Dicke's superradiance (SR) and first described in \citet{Dicke1954}. In the SR process a sample of excited molecules characterised by slow relaxation and dephasing time scales evolves, through interaction with the common radiation field, into a quantum state possessing a high degree of entanglement between the constituent molecules' individual excitation states. These entangled states couple strongly to the radiation field and produce enhanced emission rates as compared to those rates predicted for independently radiating molecules. Recent work \citep{Rajabi2016B,Rajabi2017,Houde2018a,Houde2019,Rajabi2019,Rajabi2020,Rajabi2020b} has demonstrated that SR is a strong candidate for describing transient astrophysical processes demonstrating sharp flux rises in maser-harbouring regions, which possess conditions similar to those prerequisite to the development of SR. In \cite{Rajabi2017}, for example, it was argued that the relative delay ($\sim \!\! 20$ days) and differing duration ($\sim \!\! 7$ days and $\sim \!\! 20$ days) of similarly periodic ($\sim \!\! 34.4$ days) methanol 6.7-GHz and water 22-GHz flares (respectively), observed in the intermediate-mass young stellar object G107.298+5.639 \citep{Szymczak2016}, is difficult to explain by the quasi-steady state dynamics of a maser model with periodic pumping source. Conversely, an SR numerical model naturally reproduced the distinct temporal timescales between both flares, while being triggered by a single common periodic population inversion source.

The recent applications of SR to astrophysics \citep{Rajabi2016B,Rajabi2017,Houde2018a,Houde2019,Rajabi2019,Rajabi2020,Rajabi2020b} have been restricted to slices of velocity coherent (on resonance) populations, where the relevant Maxwell-Bloch (MB) equations describe a gas of molecules travelling at only a single shared velocity. It is the objective of the present work to extend modelling of SR to realistic velocity distributions. Such an extension is ultimately motivated by our desire to eventually study coherence in observational data. 

The emergence of coherence in the transfer of radiation through stimulated emission processes has a long history in the theory of astrophysical maser propagation \citep{Elitzur1992,Gray2012}. It is commonly assumed that only processes of absorption and stimulated emission contribute to the propagation of maser radiation \citep{Goldreich1972,Elitzur1992,Gray2012}. Such modelling is justified in the quasi-steady state limit \citep{Rajabi2020}, under the assumptions of (i) incoherent radiation, where the radiation field frequency modes fulfill Gaussian statistics and are uncorrelated \citep{Litvak1970}, and (ii) the steady state of the molecular populations \citep{Litvak1970,Goldreich1972,Gray2012}. Proper modelling of the amplification of radiation in a population inverted medium has revealed that these assumptions are warranted for unsaturated masers \citep{Trung2009a,Trung2009b}, where the dephasing timescale is much shorter than the timescale of stimulated emission processes.

However, already at low degrees of maser saturation it has been shown that coherence emerges in both the radiation field and between the populations \citep{Menegozzi1978,Trung2009a,Trung2009b}, rendering the quasi-steady state limit invalid \citep{Gray2012}. Attempts have been made to partially account for the coherence properties of radiation \citep{Field1984,Field1988}, but their utility is limited to low degrees of saturation. To properly account for the coherence properties in the maser process, one needs to solve the full MB equations across a velocity distribution \citep{Gray2012,Menegozzi1978, Sargent1974}, as was done in \citet{Menegozzi1978} and previously in quantum optics studies of SR \citep{Sargent1974, MacGillivray1976, Benedict1996}. In light of the efficiency and effectiveness of the methods of \citet{Menegozzi1978} in describing the emergence of coherence in the maser regime of the MB equations, we investigate in this paper the application of their methods to the high field transient SR regime.

The paper is organised as follows. In Section \ref{sec:collective_emission_and_MnL} we introduce methods for simulating collective and cooperative emission processes across velocity distributions and we derive the time domain envelope factorisation of the MB equations for modelling a one-dimensional quasi-steady state maser process and a transient SR process. After discussing the computational complexity of the time domain representation, the Fourier space representation of \citet{Menegozzi1978} is introduced (herein referred to as the ML representation and its solution method as the ML algorithm). We take the opportunity at this point to generalise an approximation method of \citet{Menegozzi1978} to the so-called local mode interaction (LMI) approximation, which offers higher degrees of fidelity where regions of higher field strength demand and which translates in a straightforward manner to the transient work of later sections. The ML method's advantageous computational scaling complexity and drawbacks compared to the time domain method is discussed.

In Section \ref{sec:transient_vs_ss} we describe the fundamental distinctions between maser and SR processes which we expect to complicate the application of the ML algorithm to SR. We discuss the validity of a periodic temporal Fourier series representation of the spatial propagation of spectral noise in a quasi-steady state maser process (as conducted by \citealt{Menegozzi1978}) and we discuss the challenges faced by such a representation when simulating a transient SR process.

In Section \ref{sec:ML_perf} we investigate the performance of the ML algorithm in the transition from the unsaturated maser to the saturated maser quasi-steady states, as well as in the high field transient SR regime. We simulate a one-dimensional sample configured to demonstrate all such processes at different positions along its length. We make two evaluations of the ML algorithm with this system. First, we investigate the performance of the LMI approximation in the transition from regions characterised by weak field (unsaturated) masers to regions characterised by strong field (saturated) masers. Second, after making a minor revision which enables us to enforce temporal initial conditions, we investigate the ML algorithm's ability to model transient SR processes within the sample.

Upon demonstrating the inaccuracy of the transient application of the ML algorithm, we proceed to the central work of this paper. In Section \ref{sec:if_MBEs} we construct a manifestly unique Fourier representation of the MB equations which generalises the ML algorithm. This representation is capable of modelling high field strength transient SR processes, may be executed with varying degrees of approximation fidelity, and retains the conventional ML algorithm's improved computational complexity scaling over the time domain method. We demonstrate the successful simulation of all SR regions of the system investigated in the prior Section \ref{sec:ML_perf}, and we characterise approximation fidelity requirements for future simulations of transient SR processes with our new algorithm.

A list of abbreviations is provided in Appendix \ref{app:abbr}. Appendix \ref{app:lmi_just} provides a rigorous justification for the LMI approximation from perturbation theory. Appendix \ref{app:if_mbes_re_n_im} expresses the novel Fourier representation central to this paper in a format more naturally suited to numerical simulation; namely, in its real and imaginary parts.

\section{Collective and cooperative emission processes across velocity distributions and the Menegozzi \& Lamb Method}\label{sec:collective_emission_and_MnL}

\subsection{Modelling maser and superradiant processes across velocity distributions}\label{subsec:coherent_within_incoherent}

There are three common methods for modelling the maser action across wide incoherent velocity distributions: first, by a theory of rate-balanced excitation and de-excitation of velocity sub-populations, with accompanying equations of radiative transfer \citep{Elitzur1992}; second, by the master equation describing the evolution of the quantum mechanical density operator \citep{Goldreich1974, Menegozzi1978, Gray2012}; and third, by the Heisenberg equations describing the time evolution of expectation values of the population inversion, polarisation, and field operators within the Heisenberg picture of QED \citep{Gross1982, Rajabi2016A}. The latter two methods lead to the velocity dependent MB equations. All three methods must model, in some manner, the relatively weak coherence of the maser action within narrow velocity slices of the global incoherent velocity distribution. Compared to the intensity that would be generated by a fully coherent population sharing a single velocity, the total intensity generated by the incoherent distribution is reduced by the independence of these velocity slices.

Maser rate-balancing algorithms treat the radiators as statistically independent, and therefore do not generalise to SR modelling; conversely, the MB equations--being derived from the fully quantum mechanical density operator master equation or Heisenberg equations--continue to describe cooperative coherent emission in the transient SR regime. The MB equations are the starting point of \citet{Menegozzi1978} and of our present work. A derivation of the MB equations as a valid representation of transient SR processes (under reasonable approximations) can be found in the literature \citep{Arecchi1970, MacGillivray1976, Gross1982, Andreev1990, Benedict1996, Rajabi2016A}. We turn now to discuss the MB equations over a velocity distribution, with the objective of constructing a numerically efficient algorithm for solving them in the transient SR regime.

\subsection{The Maxwell-Bloch equations and the slowly-varying envelope approximation}\label{subsec:env_fact_of_MBes}

Derivations of the MB equations start from the two-level (with ground state $|g\rangle$ and excited state $|e\rangle$) model of a molecule possessing an electric or magnetic transition matrix element. The population inversion density $N\left(\mathbf{r}\right)$ of the sample is defined as a coarse-grained function over the sample volume, valued at the $k^\text{th}$ molecular site with the expectation value of the $k^\text{th}$ molecule's population inversion operator $|e_k\rangle \langle e_k|-|g_k\rangle \langle g_k|$. The polarisation $\mathbf{P}\left(\mathbf{r}\right)$ is a coarse-grained vector field over the sample volume, valued at each site with the expectation value of the molecular dipole operator weighted by the local population density.\footnote{In the case of the electric dipole transition in the two-level basis, the dipole operator of the $k^\text{th}$ molecule is $\mathbf{d}\left(|e_k\rangle \langle g_k|+|g_k\rangle \langle e_k|\right)$ for a molecule with dipole moment $\mathbf{d}$.} The quantum mechanical Heisenberg equations determine the self-consistent evolution of the population inversion density, the polarisation or magnetisation, and the field amplitude through the MB equations \citep{Arecchi1970, MacGillivray1976, Gross1982, Benedict1996, Rajabi2016A, Rajabi2016B}.

For a one-dimensional sample extended along the $z$ axis with all dipole moments, the media polarisation, and the field polarisation oriented along a fixed orientation perpendicular to the $z$ axis, the MB equations across a velocity distribution with an electric dipole transition are \citep{Gross1982}
\begin{align}
    \left[\frac{\partial}{\partial t} + v \frac{\partial}{\partial z}\right]N_{v} &= \frac{i}{\hbar} \left(E^{+}+E^{-}\right) \left(P_{v}^{+} - P_{v}^{-}\right) \label{eq:MBE_Inv} \\
    \left[\frac{\partial}{\partial t} + v \frac{\partial}{\partial z} \right] P_{v}^{+} &= i \omega_{0} P_{v}^{+} + 2i \frac{d^{2}}{\hbar} \left(E^{+}+E^{-}\right) N_{v} \label{eq:MBE_Pol} \\
    \left[\frac{\partial^{2}}{\partial t^{2}} - c^{2} \frac{\partial^{2}}{\partial z^{2}}\right] E^{+} &= -\frac{1}{\epsilon_{0}} \int \mathrm{d} v F(v) \frac{\partial^2 P_{v}^{-}}{\partial t^2}, \label{eq:MBE_Field}
\end{align}
where $N_{v}$ is half the population inversion for those molecules travelling with velocity $v$, $P_{v}^{\pm}$ are the forward ($+$) and reverse ($-$) rotating parts\footnote{If $f\left(z,t\right)=\int_{-\infty}^{+\infty}\tilde{f}\left(z,\omega\right)e^{i \omega t} \mathrm{d}\omega$, then $f^{\pm}\left(z,t\right)$ are defined as $f^{-}\left(z,t\right)=\int_{-\infty}^{0}\tilde{f}\left(z,\omega\right)e^{i \omega t} \mathrm{d}\omega$ and $f^{+}\left(z,t\right)=\int_{0}^{+\infty}\tilde{f}\left(z,\omega\right)e^{i \omega t} \mathrm{d}\omega$.} of the polarisation for those molecules travelling with velocity $v$, and $E^{\pm}$ are the forward and reverse rotating parts of the electric field. We note that in the quantum mechanical limit from which the derivation of the MB equations starts, the $P^+$ ($P^-$) correspond to molecular raising (lowering) operators and the $E^+$ ($E^-$) to photon annihilation (creation) operators. All quantities depend upon only position $z$ and time $t$. The angular frequency of emission is $\omega_{0}$ in the rest frame, the molecular dipole moment is $d$, and $F\left(v\right)$ is defined such that the fraction of molecules of velocity between $v$ and $v+dv$ is $F\left(v\right)dv$ (where $\int F\left(v\right) dv = 1$).

We now make a change of variables to the retarded time $\tau=t-z/c$ and factor the polarisations and electric field with envelope functions\footnote{The ``bar'' on $\bar{\mathcal{P}}_{v}$ distinguishes our polarisation envelopes from the literature, in that we factor by Doppler shifted frequencies on a per-channel basis.} as
\begin{align}
    P_{v}^{\pm}\left(z,\tau\right) & =\bar{\mathcal{P}}_{v}^{\pm}\left(z,\tau\right)e^{\pm i\omega_{0}\left(1+v/c\right)\tau}\\
    E^{\pm}\left(z,\tau\right) & =\mathcal{E}^{\pm}\left(z,\tau\right)e^{\mp i\omega_{0}\tau}\label{eq:Efactorisation}.
\end{align}
If we neglect the fast-rotating terms $E^+ P^-_v$ and $E^- P^+_v$ (the so-called ``rotating wave approximation'') in equation (\ref{eq:MBE_Inv}) and recognise that $\partial/\partial z\ll\omega_{0}/c$ and $\partial/\partial \tau \ll \omega_0$ when acting on the envelope functions, we arrive at the so-called slowly-varying envelope approximation (SVEA) of the MB equations \citep{Gross1982},\footnote{Our form of the SVEA of the MB equations with a velocity distribution differs slightly from that of \citet{Gross1982} or \citet{Andreev1990} due to our Doppler shifting of the polarisation envelopes.}
\begin{align}
    \frac{\partial N_{v}}{\partial\tau} &= \frac{i}{\hbar} \left(\bar{\mathcal{P}}_{v}^{+}\mathcal{E}^{+}e^{i\omega_{0}\frac{v}{c}\tau} - \bar{\mathcal{P}}_{v}^{-}\mathcal{E}^{-}e^{-i\omega_{0}\frac{v}{c}\tau}\right) - \frac{N_{v}}{T_{1}} + \Lambda^{(N)} \label{eq:MB_TD-1} \\
    \frac{\partial\bar{\mathcal{P}}_{v}^{+}}{\partial\tau} &= i \frac{2d^{2}}{\hbar} \mathcal{E}^{-}N_{v}e^{-i\omega_{0}\frac{v}{c}\tau} - \frac{\bar{\mathcal{P}}_{v}^{+}}{T_{2}} + \Lambda^{(P)} \label{eq:MB_TD-2} \\
    \frac{\partial\mathcal{E}^{+}}{\partial z} &= i \frac{\omega_{0}}{2\epsilon_{0}c} \int \mathrm{d} v \left(1+\frac{v}{c}\right) F(v) \mathcal{\bar{P}}_{v}^{-}e^{-i\omega_{0}\frac{v}{c}\tau}, \label{eq:MB_TD-3}
\end{align}
where we have introduced population inversion and polarisation pumping sources $\Lambda^{(N)}\left(\tau\right)$ and $\Lambda^{(P)}\left(\tau\right)$, respectively, as well as non-coherent relaxation and dephasing time scales $T_{1}$ and $T_{2}$, respectively. The factor $\left(1+v/c\right)$ in \eqref{eq:MB_TD-3} is retained only for the discussion of the following paragraph, but is replaced in all practical computations by $1+v/c \approx 1$. We make two observations on equations \eqref{eq:MB_TD-1}--\eqref{eq:MB_TD-3}.

First, if the velocity distribution $F\left(v\right)$ is a purely coherent one at some velocity $v_{0}$--that is, if $F\left(v\right)=\delta\left(v-v_{0}\right)$--then our choice of Doppler shifted polarisation envelope frequency causes the system of equations \eqref{eq:MB_TD-1}--\eqref{eq:MB_TD-3} to appear exactly as a coherent system with $\omega_0 \rightarrow \omega'_{0}=\omega_{0}\left(1+v_0/c\right)$, if only we redefine our field factorisation of equation \eqref{eq:Efactorisation} by the Doppler shifted frequency $\omega'_0$ (upon doing so all manifest velocity reference vanishes and all occurrences of $\omega_0$ become $\omega'_0$). Therefore, although equations \eqref{eq:MB_TD-1}--\eqref{eq:MB_TD-3} are not manifestly symmetric across velocity channels, the channel dependent factors of $v$ are merely artifacts of our choice of reference electric field envelope frequency $\omega_0$. Each velocity slice sees a physically equivalent system centred upon its own Doppler shifted natural frequency of oscillation. When generalising to a wide velocity distribution $F\left(v\right)\neq\delta\left(v-v_{0}\right)$, a given velocity slice should couple most strongly to those Fourier components of the field neighbouring its natural Doppler shifted frequency. This physical argument will motivate the local mode interaction approximation in Section \ref{subsubsec:lmia}.

Second, because the physics should be symmetric across velocities, we decay and pump the polarisation on resonance with a velocity channel's Doppler shifted frequency; i.e., no velocity dependent exponential multiplies $\bar{\mathcal{P}}_{v}^{\pm}/T_{2}$ nor $\Lambda^{(P)}$ in equation (\ref{eq:MB_TD-2}), despite $\bar{\mathcal{P}}_{v}$ being Doppler shifted relative to $P_{v}^{\pm}$ of equation (\ref{eq:MBE_Pol}). Had we originally introduced pumping and decay terms to equation (\ref{eq:MBE_Pol}), we would have accidentally neglected this physical symmetry across velocities.

\subsection{Spontaneous emission and the initial Bloch angle prescription}\label{subsec:sp_em_n_init_bloch}

We stated in Section \ref{subsec:coherent_within_incoherent}, without proof, a basic tenet of this paper: that the MB equations accurately model SR transient processes. A precise derivation of this result can be found in the literature \citep{Arecchi1970, MacGillivray1976, Gross1982, Benedict1996}, but it is necessary to describe here a feature of the derivation relevant to the modelling of transient processes with the MB equations.

It is apparent from equations \eqref{eq:MBE_Inv}--\eqref{eq:MBE_Field} that the MB equations alone will not model even the simplest spontaneous emission process. Starting from an initially inverted population possessing no polarisation, and in the absence of an electric field, we expect a sample to spontaneously emit photons and eventually generate a non-zero electric field and non-zero polarisation. Instead, upon inspection of the SVEA MB equations, we see that the null right side of equation \eqref{eq:MB_TD-2} will never allow the sample to acquire a polarisation (nor an electric field).

In order to model any spontaneous emission processes, the polarisation initial conditions of the semi-classical MB equations must be prescribed by a purely quantum mechanical analysis. In \citet{Gross1982} it is shown that interaction of the inverted molecules with fluctuations of the quantized radiation field leads, very quickly, to a classical ensemble of non-zero polarisation configurations. The continued evolution of the system is then described by the collection of trajectories of the MB equation determined by this ensemble of initial conditions. Transients of population inversion, polarisation, and fields are computed from expectation values averaged over these simulated trajectories.

In fact, the averaging operation yields negligible modifications to our results. We will be concerned only with the degree of polarisation built up in the interaction of the initial population inversion with fluctuations of the quantized vacuum radiation field. This polarisation value is \citep{Polder1979,Gross1982} that for which the conventional Bloch angle $\theta_{\text{B}}$ (defined via $\tan\left[\theta_\mathrm{B}\left(z,\tau\right)\right] = \left|P \left(z,\tau\right)\right| / \left[d N\left(z,\tau\right)\right]$) has tipped to $\theta_{\text{B},0}=2/\sqrt{N_{\text{mol}}}$, where $N_{\text{mol}}$ is the number of molecules in the sample.\footnote{Actually, $N_{\text{mol}}$ should be replaced here by the number of \emph{interacting} molecules. This is problematic, as the number of interacting molecules within a velocity distribution is not well-defined at this point in our analysis. We discuss this point further in Section \ref{subsec:future_work} on future research.} For a sample of a large number of molecules with initial population inversion $N_0$, this prescribes an initial polarisation $P_0$ according to $P_0 / d = N_{0} \tan\left(\theta_{\text{B},0}\right) \approx N_{0} 2 / \sqrt{N_{\text{mol}}}$.

\subsection{Computational complexity of the Maxwell-Bloch equations in the time domain}\label{subsec:td_complexity}

Although equations \eqref{eq:MB_TD-1}--\eqref{eq:MB_TD-3} analytically remove the stiffest\footnote{The term "stiff" has various usages in the literature. A "stiff" term in this paper is any derivative generating term (any term on the right side of our differential equations as written) which places finer step size demands (relative to adjacent generating terms) upon the numerical algorithm.} temporal propagation term $i\omega_{0}P_{v}^{+}$ from equation (\ref{eq:MBE_Pol}), a lower degree of stiffness remains present within the exponentials $\exp\left[\pm i\omega_{0}\left(v/c\right)\tau\right]$ (had we not Doppler shifted our polarisation envelopes, this stiffness would have emerged in a term of the form $i\omega_{0}\left(v/c\right)\mathcal{P}_{v}^{+}$).

The order of total numerical complexity to a Runge-Kutta solution of equations \eqref{eq:MB_TD-1}--\eqref{eq:MB_TD-3} is degraded by the increasing stiffness of these exponential terms with increasing width of the velocity distribution. As $F\left(v\right)$ widens, these exponentials oscillate at higher frequencies and demand finer time stepping to avoid aliasing of their cycles. No analytical factorisation can remove this stiffness, which forces a time domain algorithm to be $\mathcal{O}\left(N^{2}\right)$ complex in the number $N$ of velocity channels simulated. Doubling the velocity width, for example, demands both that twice as many channels be simulated and that each be simulated with twice as fine a time step in a Runge-Kutta propagation of equations \eqref{eq:MB_TD-1} and \eqref{eq:MB_TD-2}.

\subsection{The Menegozzi \& Lamb Method}\label{subsec:ML_alg}

We desire to reduce the order of numerical complexity in simulating our system by turning to physical arguments. To this end, we review in this section the one-dimensional maser simulation algorithm developed by \citet{Menegozzi1978} within a temporal Fourier series representation of the MB equations.

This representation will introduce two numerical advantages. First, it will allow a simulation to crop the spectrum of each velocity channel's population inversion and polarisation transients to those spectral components lying within a limited neighbourhood of the channel's natural Doppler shifted frequency. Second, it will enable the assertion of what we refer to here as the LMI approximation, which suppresses the algebraic coupling of inversion and polarisation velocity channels to electric field modes sufficiently far removed from their natural frequencies. This approximation is presented in only a limiting case in \citet{Menegozzi1978}, but is naturally generalised in the present work.

Importantly, although the LMI approximation will remove the formal direct mathematical coupling between distant frequency modes, it will not necessarily remove the possibility of physical coupling and correlation between distant modes through indirect, transitive\footnote{We colloquially describe ``transitive'' coupling by the following example. Suppose that velocity channels $A,\:B,\:C$ would naturally radiate field modes of frequencies $\omega_{A}<\omega_{B}<\omega_{C}$, and that our LMI approximation is only so sufficiently wide as to mathematically couple $A\rightleftharpoons\left\{ \omega_{A},\omega_{B}\right\} ;\:B\rightleftharpoons\left\{ \omega_{A},\omega_{B},\omega_{C}\right\} ;\:C\rightleftharpoons\left\{ \omega_{B},\omega_{C}\right\}$ (``$\rightleftharpoons$'' denotes ``couples to''). Although channel A does not \emph{mathematically} couple to field mode $\omega_{C}$, we recognise that channel $A$ may indirectly \emph{physically} couple to $\omega_{C}$ via the transitive coupling $A\rightleftharpoons\omega_{B}\rightleftharpoons B\rightleftharpoons\omega_{C}$.} means. In Section \ref{subsubsec:lmia_ifr} we will investigate the accuracy of this approximation in the SR domain; it will then become the task of future research, operating within these LMI approximation fidelity constraints, to quantify the degree of transitive coupling and correlation between SR processes across a broad velocity distribution. The algorithm we develop in Section \ref{sec:if_MBEs} will enable this future research.

\vspace{12pt}\noindent In the remainder of this Section \ref{subsec:ML_alg} the notation is our own but the theory loosely follows that of \citet{Menegozzi1978}.

\subsubsection{The Menegozzi \& Lamb representation of the Maxwell-Bloch equations}\label{subsubsec:MLr_MBEs}

For a simulation of duration $T$, the population inversion, the polarisation envelopes, the pumping sources, and the field envelopes are expanded in Fourier series of mode separation $d\omega=2\pi/T$. Additionally, the velocity distribution is partitioned with a granularity $dv$ of the equivalent Doppler shift $d\omega$ between adjacent channels; namely, $dv=cd\omega/\omega_{0}$. If $p$ denotes the integer multiple of $dv$ identifying a velocity channel then, for example, $\mathbb{N}_{p,m}$ denotes the $m^{\text{th}}$ frequency mode of the population inversion of a velocity slice centred at velocity $v=pdv$. Explicitly, the expansions read as
\begin{align}
    N_{p} &= \sum_{m} \mathbb{N}_{p,m} \left(z\right) e^{i m d\!\omega\tau} \\
    \mathcal{\bar{P}}_{p}^{\pm} &= \sum_{m} \bar{\mathbb{P}}_{p,m}^{\pm} \left(z\right) e^{\pm i m d\!\omega\tau} \\
    \mathcal{E}^{\pm} &=\sum_{m} \mathbb{E}_{m}^{\pm} \left(z\right) e^{\mp i m d\!\omega\tau} \\
    \Lambda^{\left(N/P\right)} &= \sum_{m} \mathbb{L}^{\left(N/P\right)}_m e^{i m d\!\omega\tau}.
\end{align}
Upon substitution into the MB equations and some changes of summation orders and variables (the details of which are omitted here), the first two MB equations (\ref{eq:MB_TD-1}) and (\ref{eq:MB_TD-2}) translate to the algebraic Fourier mode relations
\begin{align}
    \begin{split}
        \left(i m d\omega \right) \mathbb{N}_{p,m} &= \frac{i}{\hbar} \sum_{\bar{m}} \left(\bar{\mathbb{P}}_{p,\bar{m}}^{+} \mathbb{E}_{p+\bar{m}-m}^{+} - \bar{\mathbb{P}}_{p,\bar{m}}^{-} \mathbb{E}_{p+\bar{m}-m}^{-}\right) \\
        &\quad -\frac{\mathbb{N}_{p,m}}{T_{1}} + \mathbb{L}_{m}^{(N)} \label{eq:FourierNpm}
    \end{split} \\
    \begin{split}
        \left(i m d \omega \right) \bar{\mathbb{P}}_{p,m}^{+} &= \frac{ i 2 d^{2}}{\hbar} \sum_{\bar{m}} \left(\mathbb{N}_{p,\bar{m}} \mathbb{E}_{p+m-\bar{m}}^{-}\right) \\
        &\quad -\frac{\bar{\mathbb{P}}_{p,m}^{+}}{T_{2}} + \mathbb{L}_{m}^{(P)}. \label{eq:FourierPpm}
    \end{split}
\end{align}
Equations (\ref{eq:FourierNpm}) and (\ref{eq:FourierPpm}) are mathematically equivalent to equations (2.30) and (2.29), respectively, of \citet{Menegozzi1978}, despite differing notation and algebraic rearrangement. The form here will prove advantageous for our work generalising the algorithm to the transient domain in Section \ref{sec:if_MBEs} and for our introduction of the LMI approximation. The third MB equation \eqref{eq:MB_TD-3} reads
\begin{equation}
    \frac{\partial\mathbb{E}_{m}^{+}}{\partial z} = i \frac{d\omega}{2\epsilon_{0}} \sum_{p} F_{p} \bar{\mathbb{P}}_{p,m-p}^{-}, \label{eq:zPropFour}
\end{equation}
where $F_{p}=F\left(pdv\right)$. We refer to equations \eqref{eq:FourierNpm}--\eqref{eq:zPropFour} as the Menegozzi \& Lamb (ML) representation of the MB equations.

\subsubsection{Solution method, spectral limiting, and the local mode interaction approximation}\label{subsubsec:lmia}

Solving the ML equations is a straightforward numerical task. Starting from the Fourier representation $\mathbb{E}_{m}^{\pm}$ of a given incident electric field time dependence $E\left(z=0,\tau\right)$ at the start of the sample, equations (\ref{eq:FourierNpm}) and (\ref{eq:FourierPpm}) form a linear system, which is solved for the population inversion and polarisation modes at $z=0$; next, the polarisation modes are used to propagate the electric field modes forward one step in $z$ via equation (\ref{eq:zPropFour}). These two steps loop along the entire length of the sample. In a practical numerical scheme, a fourth-order Runge-Kutta abstraction of the $z$-stepping is employed.

In their native form, equations \eqref{eq:FourierNpm} and \eqref{eq:FourierPpm} are computationally expensive in the number of velocity channels $N$: for each channel added to the system, the resulting expansion of the electric field spectrum which enters on the right side of equations \eqref{eq:FourierNpm} and \eqref{eq:FourierPpm} implies expansion of the range of $m$ in $\mathbb{N}_{p,m}$ and $\bar{\mathbb{P}}^\pm_{p,m}$. Increasing the number of velocity channels thus increases the size of the linear system of unknown modes that must be solved for every velocity channel. The operation of solving a linear system of equations is $\sim\!\mathcal{O}\left(N^{3}\right)$ complex,\footnote{This complexity can be reduced to $\sim\!\mathcal{O}\left(N^{2.5}\right)$ by employing an efficient linear system solver.} so that the total algorithm is $\sim\!\mathcal{O}\left(N^{4}\right)$ complex.

The order of complexity of the ML algorithm is dramatically reduced by limiting the range of the spectral mode index of all velocity channels' population inversions and polarisations (the index $m$ of $\mathbb{N}_{p,m}$ and $\bar{\mathbb{P}}^\pm_{p,m}$). We herein refer to this approximation as \emph{spectral limiting}. If $m$ is limited to a fixed-size (independent of the number of velocity channels introduced) neighbourhood of $0$, the size of the linear system of equations for each channel does not grow with $N$; as $F\left(v\right)$ widens, numerical operations therefore increase proportional only to the number of velocity channels needing to be solved. Simulating the ML equations under spectral limiting is thus $\mathcal{O}\left(N\right)$ complex.

An additional numerical approximation introduced in Appendix C of \citet{Menegozzi1978} is to couple the population inversion and polarisation modes of a particular velocity channel to interact only with that field mode corresponding to its natural Doppler shifted resonance. In their work, \citet{Menegozzi1978} eliminate reference to the polarisation modes and achieve said approximation by recognising the dominant terms in the remaining system of equations for the unknown population inversion modes.\footnote{We refer the reader to the paragraphs immediately preceding and following equations (C11) and (C12) of \citet{Menegozzi1978}.} In our present form, and with $\bar{\mathcal{P}}_{v}$ factored about its natural Doppler shifted frequency, such an approximation is achieved by truncating to $\bar{m}=0$. This choice of term may appear at first glance ambiguous; however, a rigorous justification for it may be found in Appendix \ref{app:lmi_just}.

The formulation of the approximation in \citet{Menegozzi1978} corresponds, in our present form, to summation over the trivial set $\bar{m}\in\left\{0 \right\}$. Such an approximation is sufficiently accurate for the unsaturated maser domain but, as Menegozzi \& Lamb rightly argue, becomes inaccurate in regions of high field strength. Our formulation suggests a natural generalisation of the approximation which permits its assertion to varying degrees of fidelity. We extend the $\bar{m}$ summation over a finite neighbourhood of $0$. This is the mathematical expression of the LMI approximation introduced colloquially near the end of Section \ref{subsec:env_fact_of_MBes}. We will investigate the performance of the LMI approximation within regions of increasing field strength in Section \ref{subsec:ss_MLr}. Notice that decreasing LMI approximation fidelity (narrowing the range of the $\bar{m}$ summation) improves the sparsity of the linear system of equations \eqref{eq:FourierNpm} and \eqref{eq:FourierPpm}, and therefore offers a further reduction in numerical operations. The simulations throughout this work implement an LMI approximation of fixed fidelity across all $z$ positions. Such fidelity could, in theory, be made to vary as a function of $z$, and we discuss this possible generalisation in Section \ref{subsec:future_work} on future work.

\section{Transient superradiance processes versus quasi-steady state maser processes}\label{sec:transient_vs_ss}

In this section we differentiate between the transient nature of an SR process and the quasi-steady state nature of a maser process. For a comprehensive comparison of the two processes, see \citet{Rajabi2020}.

\subsection{Superradiance as a transient process}\label{subsec:Transient-superradiant-events}

Superradiance is fundamentally a transient phenomenon, involving a series of distinct events and the evolution from an initial energetic inverted population level to a dramatically altered final inversion level \citep{Rajabi2020}. This transient system evolution is perhaps best qualitatively understood in the Schr\"{o}dinger picture of the QED of a collection of $n$ molecules.

We imagine a system initially prepared with all molecules excited and the quantized radiation field in the vacuum state,
\begin{equation}
    |\Psi\rangle_{\text{initial}}=|e_{1}e_{2}\dots e_{n}\rangle\otimes|0\rangle_{\text{rad}},\label{eq:SPic_Ket_Init-1}
\end{equation}
where $|e_{1}e_{2}\dots e_{n}\rangle$ denotes a tensor product of molecular states which may each assume either the excited state $|e\rangle$ or the ground state $|g\rangle$, and where $|0\rangle_{\text{rad}}$ denotes the vacuum state of the radiation field.

It should be understood that the instantaneous inversion of the entire sample described by equation \eqref{eq:SPic_Ket_Init-1} is an idealisation intended, for the purpose of this section, to capture the salient features of SR. In a realistic astrophysical environment we could imagine a system starting for $t<t_\textrm{init}$ in a highly inverted state accurately modelled by maser theory. The initial preparation of equation \eqref{eq:SPic_Ket_Init-1} would be realised by a pumping flare of finite duration shorter than that of the SR transient process, but of sufficient amplitude to raise the inverted population column density above the SR threshold. In \cite{Rajabi2019}, for example, the 6.7 GHz methanol emission is believed to be pumped by infrared radiation from dust \citep{Sobolev1997} demonstrating outbursts \citep{Caratti2017, Szymczak2018b} which are proposed to initiate the onset of SR.

We assume that the relevant photon wavelengths are much larger than the size of a molecule, and we neglect the Hilbert space associated with the motions of the molecules' centres of masses. We also assume that the molecules are sufficiently separated so as not to require wavefunction symmetrisation.\footnote{Incidentally, as we describe later in this section, the calculation of \citet{Dicke1954} yields transitions through totally symmetric states only. This result is a consequence of the permutation symmetry of the interaction Hamiltonian, and not a consequence of spin-statistic imposed symmetrisation.} The total Hamiltonian $\mathcal{H}_{\text{tot}}$ of the system is the sum of the internal Hamiltonian $\mathcal{H}_{\text{mol}}$ of two-level molecular excitations, of the radiation Hamiltonian $\mathcal{H}_{\text{rad}}$, and of the field-molecule interaction Hamiltonian $\mathcal{H}_{\text{int}}$. Tensor products of free molecular excitation states and radiation field Fock states, such as that of equation (\ref{eq:SPic_Ket_Init-1}), are eigenstates of $\mathcal{H}_{0}\equiv\mathcal{H}_{\text{mol}}+\mathcal{H}_{\text{rad}}$, but not of $\mathcal{H}_{\text{tot}}=\mathcal{H}_{0}+\mathcal{H}_{\text{int}}$. Treating $\mathcal{H}_{\text{int}}$ as a perturbation to $\mathcal{H}_{0}$ yields transition amplitudes between eigenstates of $\mathcal{H}_{0}$.

In his seminal paper \citet{Dicke1954} describes first the small sample limit, defined such that molecules are separated by a distance much less than their spontaneous emission wavelength $\lambda$ but much greater than any intermolecular interaction length. Although the molecules are fundamentally distinguishable in the small sample limit, they are not distinguishable via observation of an emitted photon. Thus, if spontaneous emission is described by the transition to some singular de-excited molecular state $|\phi\rangle_{\text{mol}}$ with an accompanying photon of mode $p$,
\begin{equation}
    |e_{1}e_{2}\dots e_{n}\rangle_{\text{mol}} \otimes |0\rangle_{\text{rad}} \rightarrow |\phi\rangle_{\text{mol}} \otimes |1_{p}\rangle_{\text{rad}}, \label{eq:SR_emission}
\end{equation}
then the final molecular state $|\phi\rangle_{\text{mol}}$ must be indeterminate in the identity of the ground state molecule. The precise calculations of \citet{Dicke1954} show that the most  probable molecular state is the symmetric superposition of all possible configurations having one molecule in the ground state; i.e., $|\phi\rangle_{\text{mol}}=\left(1/\sqrt{n}\right)\sum_{k}|e_{1}e_{2}\dots g_{k}\dots e_{n}\rangle\equiv|s\left(1\right)\rangle$, where we define $|s\left(k\right)\rangle$ as the symmetric superposition of all states possessing $k$ molecules in the ground state. Note that $|s\left(k\right)\rangle$ is an entangled state for $0<k<n$. By energy conservation, the energy of the emitted photon matches the loss in molecular excitation energy $\hbar\omega_0$.

\begin{figure}
    \begin{align*}
        |s\left(0\right)\rangle &= |e_1 e_2 \dots e_n\rangle  \\
        &\Big\downarrow \rightsquigarrow \hbar \omega_{0} \\
        |s\left(1\right)\rangle &= \frac{1}{\sqrt{n}} \left( |g_1 e_2 \dots e_n\rangle + |e_1 g_2 \dots e_n\rangle + \dots + |e_1 e_2 \dots g_n\rangle \right) \\
        &\Big\downarrow \rightsquigarrow \hbar \omega_{0} \\
        &\!\!\!\dots \\
        &\Big\downarrow \rightsquigarrow \hbar \omega_{0} \\
        |s\left(n\right)\rangle &= |g_1 g_2 \dots g_n \rangle
    \end{align*}
    
    \caption{Schematic of the cascade down the ladder of symmetric excitation states with accompanying photon emission.}
    \label{fig:Cascade}
\end{figure}

The transient SR process is the cumulative effect of the cascade down the totally symmetric excitation states $|s\left(k\right)\rangle$ with accompanying photon emission as depicted in Figure \ref{fig:Cascade}, where emission rates vary with $k$. The full perturbation calculation finds that the halfway state $|s\left(n/2\right)\rangle$\footnote{for even $n$; $|s\left(n/2 \pm 1/2\right)\rangle$ for odd $n$} couples most strongly to the radiation field, so that emission is maximised after some delay $\tau_\text{D}$ required to reach this state. The cascade may be modelled by a Markovian traversal through the symmetric states with a Lindblad operator describing photon loss to the environment. Analysis of transitions between symmetric states provides a radiation intensity transient, where the intensity is derived from the expectation value of the transition rate as a function of time, as averaged over many repetitions of the full stochastic cascade \citep{Dicke1954, Rajabi2016A, Steck2020}.\footnote{Averaging is justified in astrophysical observations by the fact that the unresolved source is composed of many statistically independent SR cylinders, each representing a single realisation of the SR cascade experiment.}

The SR process is thus an inherently transient one, characterised by discrete events and phases: first, a pumping event initiates the inversion of the system at some time $\tau_{0}$; second, the system evolves over a time $\tau_\text{D}$ to the maximal emission state $|s\left(n/2\right)\rangle$; and third, the emission eventually concludes when the system has evolved into the final fully ground state $|s\left(n\right)\rangle=|g_{1}g_{2}\dots g_{n}\rangle$.

This qualitative evolution carries over, with some modification, to the more complicated case of an extended sample with molecules distributed over distances much greater than $\lambda$. The evolution of such a system is described by equations \eqref{eq:MB_TD-1}--\eqref{eq:MB_TD-3}. We can demonstrate the transient features of SR in the simplest case of a one-dimensional extended sample without a velocity distribution (at resonance) and without relaxation or dephasing effects,\footnote{If the relaxation time $T_1$ matches the dephasing time $T_2$, an analytical solution exists which generalises equation \eqref{eq:SineGordon}; see \citet{Rajabi2020}.} where the so-called sine-Gordon equation describes \citep{Gross1982, Rajabi2020} the evolution of the Bloch angle $\theta_\mathrm{B}$ according to
\begin{equation}
	\frac{\mathrm{d}^2 \theta_\mathrm{B}}{\mathrm{d} q^2} + \frac{1}{q} \frac{\mathrm{d} \theta_\mathrm{B}}{\mathrm{d} q} = \sin \left( \theta_\mathrm{B} \right), \label{eq:SineGordon}
\end{equation}
where $q$ is the dimensionless parameter $q = 2 \sqrt{ z \tau / L T_\mathrm{R}}$ for a sample of length $L$. The characteristic timescale $T_\mathrm{R}$ is determined by the sample's length, by the inverse of the molecules' Einstein coefficient of isolated spontaneous emission rate $\tau_\text{sp}$, by the spontaneous emission wavelength $\lambda$, and by the sample's inverted population density $n$ as
\begin{equation}
    T_\mathrm{R} = \tau_\text{sp} \frac{8 \pi}{3 n \lambda^2 L}.
\end{equation}
It is important to note that both the relaxation and dephasing timescales $T_1$ and $T_2$ of the sample must approximately exceed the characteristic timescale $T_\mathrm{R}$, if the system is to reach the highly entangled maximal emission state and thus demonstrate SR.

A plot of an SR transient generated by the sine-Gordon equation is depicted in Figure \ref{fig:SineGordon}, which clearly displays the transient SR phases of a buildup to the maximal emission state and a decay to the fully ground state. Note that in the extended one-dimensional sample, the initial emission can partially re-invert the sample downstream and lead to subsequent emissions, thereby producing the ringing effect visible in Section $C$ of Figure \ref{fig:SineGordon}.

\begin{center}
    \centering
    \begin{figure}
        \includegraphics[width=1.\columnwidth, trim=.25cm 0cm 0cm 0cm, clip]{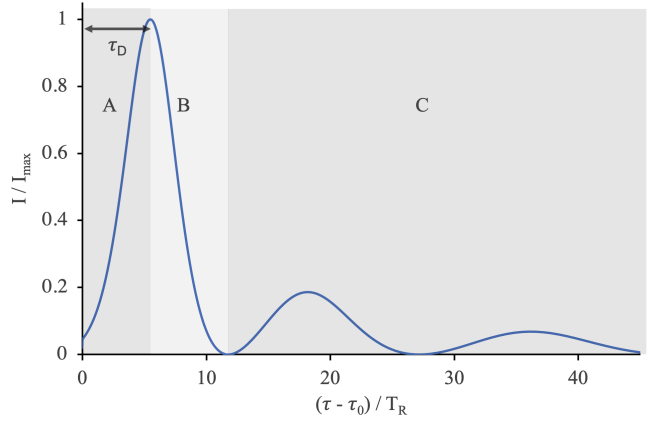}
        \caption{Normalised radiation intensity transient at the end-fire $z=L$ of a one-dimensional sample of 200 molecules, inverted at $\tau=\tau_{0}$ and exhibiting SR. Computed from the sine-Gordon limit of the MB equations. Section $A$: Buildup to the maximal emission state $|s\left(n/2\right)\rangle$. Section $B$: Decay to the fully ground state $|s\left(n\right)\rangle$. Section $C$: Re-inversion downstream and emission ringing.}
        \label{fig:SineGordon}
    \end{figure}
\end{center}

\subsection{The maser action as a quasi-steady state process}\label{subsec:ss_maser}

In contrast to SR, the maser action is effectively modelled as a quasi-steady state process involving a large number of concurrent, rate-balanced emission and stimulation events \citep{Feld1980,Elitzur1992,Gray2012,Rajabi2020}. These events are connected across molecular sites through the seed photon: a photon of mode $p$ emitted at a molecular site $j$ can enhance the probability of emission into the same mode $p$ at another molecular site $k$ through a process such as, for example,
\begin{equation}
    |e_{k}\rangle_{\text{mol}} \otimes |1_{p}\rangle_{\text{rad}} \rightarrow |g_{k}\rangle_{\text{mol}} \otimes |2_{p}\rangle_{\text{rad}}.
\end{equation}
Although the maser process physically couples distinct molecular sites $j$ and $k$, the evolution between emission and stimulation events follows classical statistics. That is, emission of photon $p$ is unambiguously associated with the transition of some single molecule $j$ from the excited to the ground state; then, under the assumption that molecule $j$ is in the ground state and that the radiation field possesses photon $p$, the probability of stimulation of a photon $p$ at another site $k$ may be computed; if emission occurs, molecule $k$ definitively transitions from $|e_{k}\rangle$ to $|g_{k}\rangle$.\footnote{We present these single-photon descriptions for the purpose of providing fundamental physical insight only. In typical numerical work, however, maser simulations operate on density matrices at exceedingly larger scales than those of single-photon processes. Most numerical maser models do not explicitly simulate the individual molecular events described here.}

This analysis of a large number of transition processes each into states well-defined in the identity of the emitting molecule is in contrast to the cascade process of SR. The SR photon emission discussion following equation (\ref{eq:SR_emission}), for example, described transition into a single-photon state which was an entangled superposition of the excitation states of different molecular sites; i.e., into the state $|s\left(1\right)\rangle$ that was indeterminate in the identity of the emitting molecule. In the maser case, the relaxation and dephasing timescales are much shorter than the time-scale for the evolution of the system. For a more detailed discussion of the role of relaxation, dephasing, and SR characteristic timescales in differentiating the maser and SR regimes, see \citet{Rajabi2020}.

Whereas the SR cascade unfolds as a transient process, the maser action may be successfully modelled as a quasi-steady state process; for a comprehensive summary of the theory of astrophysical masers, see \citet{Elitzur1992} and \citet{Gray2012}. The maser analysis in \citet{Menegozzi1978}, for example, investigates continuous emission in the quasi-steady state limit, where the quantity of interest to the simulation is the field spectrum profile along the sample's length. Such a profile results from the complicated inter-operation of pumping, decay, stimulated emission, and dephasing processes within a velocity distribution of molecules. Despite describing a quasi-steady state, the field spectrum certainly contains a rich ensemble of various off-resonance frequencies, and therefore varies significantly in time over the Fourier expansion period $T$ (even within the rotating envelope picture). The assumption of \citet{Menegozzi1978} is that the artificiality of such a representation's indefinite periodicity (in integer multiples of $T$) does not detract from its ability to describe quasi-steady state features within a single simulation period $T$, such as deviations from a Gaussian white noise spectral distribution along the sample's length.

It should be noted that although the maser action is a quasi-steady state process, it can demonstrate transient behaviour in the following limited sense. In the maser regime, calculation of the characteristic timescale $T_R$ yields a value much greater than either $T_1$ or $T_2$. As a result, the system tracks in lock-step with any transient behaviour of the inversion pump, the polarisation pump, or the incident electric field. A transient maser process is simply an immediate quasi-steady state response to variations in pump levels or incident field strengths. For a more thorough discussion see \citet{Rajabi2020}, where the maser domain is formally identified with those processes for which
\begin{equation}
    \frac{\partial N}{\partial \tau} \ll \frac{N}{T_1} \textrm{ and } \frac{\partial P^+}{\partial \tau} \ll \frac{P^+}{T_2}.
\end{equation}

Conversely, in an SR process $T_R \ll T_1,T_2$ so that the total transient response of the system possesses a finite memory. The inversion, polarisation, and electric field do not track in lock-step with the pump or incident field sources, and the system state at any given time depends upon its own history. When we distinguish SR as a transient process versus the maser action as a quasi-steady state process, we do so in this nuanced manner. An SR transient response to variations in pump or incident field sources is a complex dynamic process which may exhibit dramatically different timescales than those presented by the sources. It is formally identified in \citet{Rajabi2020} with those processes for which
\begin{equation}
    \frac{\partial N}{\partial \tau} \gg \frac{N}{T_1} \textrm{ and } \frac{\partial P^+}{\partial \tau} \gg \frac{P^+}{T_2}.
\end{equation}

\subsection{Fourier representations of quasi-steady state versus transient processes}\label{subsec:fr_ss_vs_tr}

The ML representation of the MB equations, albeit advantageous for numerical complexity purposes, introduces two complications. First, being an expansion in periodic basis functions, it forces periodicity upon the solutions. Second, it removes the ability to impose temporal initial conditions upon the inversion and polarisation. As discussed in our closing paragraph of the previous section, neither of these complications hinders the simulation of a quasi-steady state process, where periodicity is a reasonable approximation and where initial conditions are irrelevant. The ML algorithm is therefore naturally suited to the analysis of a quasi-steady state maser, including investigations of Gaussian white noise propagation down a sample's length or of radiation coherence (to which the algorithm was indeed applied in \citealt{Menegozzi1978}).

As discussed in Section \ref{subsec:Transient-superradiant-events}, SR is a transient process with distinct initial and final configurations. Such a process is fundamentally non-periodic, and quantities of interest (including total radiated energy, peak intensity, and process time scales) are strongly dependent upon initial conditions. We therefore expect difficulties to arise when applying the ML algorithm to SR processes, and we demonstrate shortly in Section \ref{subsec:tr_MLr} the inability of the ML algorithm to converge to the correct Fourier representation of non-periodic transient SR solutions to the MB equations.

These limitations for modelling transients may appear, upon first consideration, insurmountable by any Fourier series representation of the MB equations: any such representation being (by construction) periodic in the simulation duration $T$, and thus ill-suited to describing the evolution between significantly different initial and final configurations. Indeed, any Fourier series expansion of such a process will introduce ringing artifacts when inverted back to the time domain; however, important physical quantities of interest (such as total radiated energy, characteristic timescales, etc.) can be accurately described by a proper Fourier series expansion, if only the algorithm used is able to converge to it.

\section{Performance of the Menegozzi \& Lamb algorithm}\label{sec:ML_perf}

We investigate in this section the performance of the ML algorithm in the transition from modelling quasi-steady state maser processes at unsaturated field strengths, to maser processes at saturated field strengths, to transient SR processes. For this purpose we simulate a single one-dimensional sample which contains regions characterised by all the aforementioned processes. The experiment is detailed in Section \ref{subsec:experiment_desc} and a reference correct solution is computed in Section \ref{subsec:experiment_td} from the time domain representation of the MB equations \eqref{eq:MB_TD-1}--\eqref{eq:MB_TD-3}. The performance of the ML algorithm in computing the quasi-steady state of the system within regions of increasing field strength under varying degrees of LMI approximation fidelity is evaluated in Section \ref{subsec:ss_MLr}. After slight algebraic revision, the performance of the ML algorithm in simulating transient responses is evaluated in Section \ref{subsec:tr_MLr}, where we demonstrate the inability of the ML algorithm to converge at all to correct transients within regions demonstrating high field strength, transient SR processes.

\subsection{Experiment description}\label{subsec:experiment_desc}

A sample is initially prepared at $\tau=0$ in the fully inverted state, but with a molecular column density sufficient to initiate an SR transient only near the end of the sample ($z=L$) when a constant coherent incident electric field $E\left(z=0, \tau\right)=E_0=1\times{10}^{-16} \text{ V/m}$ is applied at the start of the sample\footnote{An incident field is not essential as the SR process can be initiated by a sufficient column density alone.} (the column density refers to the number of inverted molecules per unit area projected along the sample's full length; for a detailed discussion of the critical column density threshold necessary for SR, see \citealt{Rajabi2020}). The sample possesses a velocity distribution of narrow extent; specifically, $21$ velocity channels are simulated, separated by the fundamental velocity differential $dv=\left(2\pi/T\right)\left(c/\omega_0\right)$ established by the simulation duration $T$.

We simulate the cylindrical sample of methanol molecules ($\omega_0 = 2 \pi \times 6.7 \text{ GHz}$, $d=0.7\text{ D}$) described in \citet{Rajabi2020} over a duration $T={10}^8 \text{ s}$, having length $L=2 \times {10}^{15} \text{ cm}$, radius $w=5.4 \times {10}^{7} \text{ cm}$, population inversion relaxation time constant $T_1=1.64 \times {10}^{7} \text{ s}$, and polarisation dephasing time constant $T_2=1.55 \times {10}^{6} \text{ s}$. Despite such time scales being orders of magnitude longer than those typically used in maser models of star formation regions, there is in theory no physical barrier to their occurrence. In \citet{Rajabi2019} it is shown that $T_2 = 1.55 \times {10}^{6} \text{ s}$ corresponds to a gas density of approximately $10^5 \text{ cm}^{-3}$. Rather than precluding such a large value of $T_2$ on account of an assumption of high gas density, the very natural fit of the SR model in \citet{Rajabi2019} under such a value of $T_2$ may be considered evidence of low gas density in the star formation region which it models.

Our simulation differs from that of \citet{Rajabi2020} in our initial population inversion of $N_0=1.5 \times {10}^{-12} \text{ cm}^{-3}$ at $\tau=0$ (cf. $N_0=3.3\times{10}^{-12} \text{ cm}^{-3}$ in \citealt{Rajabi2020}) and in our non-vanishing incident $E_0$.\footnote{This difference being in addition to the key distinction that the present work models a (non-trivial) velocity distribution.} We apply a constant restoring population inversion pump equal to the relaxation rate; i.e., $\Lambda^{\left(N\right)} \left(\tau\right) = N_0/ \left(2 T_1\right)$.\footnote{Recall that $N_v$ is defined as half the population inversion, hence the factor of $1/2$ in the restoring pump.} The velocity distribution is uniform, so that $F\left(v\right)=1 / \Delta v$ for the total velocity width $\Delta v = 21 dv = 9.4 \times {10}^{-9} \text{ m/s}$. The molecular density for the present $\Delta v$ corresponds to an inverted molecular density on the order of $0.1 \text{ cm}^{-3}$ for a realistic velocity distribution of $\Delta v \approx 1 \text{ km/s}$. Note that the cylindrical dimensions correspond to a Fresnel number $\pi w^2 / L \lambda$ of unity.

\subsection{Reference time domain solution}\label{subsec:experiment_td}

Reference intensity transients $I\left(z,\tau\right)=c \epsilon_0 |E\left(z,\tau\right)|^2/2$ at each of six positions along the length of the sample, normalised to the incident intensity $I_0 = c \epsilon_0 |E_0|^2/2$ at the start of the sample, are readily computed from the time domain representation of the MB equations \eqref{eq:MB_TD-1}--\eqref{eq:MB_TD-3}. Our solution is a generalisation of the technique of \citet{Mathews2017}, \citet{Houde2019}, \citet{Rajabi2019}, and \citet{Rajabi2020} to a distribution of velocity channels. To commence the simulation, the inverted populations and polarisations of all velocity channels are set to their initial conditions at $\tau=0$ (recalling the polarisation initial condition prescription of Section \ref{subsec:sp_em_n_init_bloch}), and the electric field throughout the sample is initialised via a fourth-order Runge-Kutta $z$-propagation of equation \eqref{eq:MB_TD-3} from $z=0$ to $z=L$. The electric field is then used to perform a fourth-order Runge-Kutta time advancement ($\tau\rightarrow\tau+d\tau$) of all population inversion and polarisation velocity channels, the electric field is re-propagated along $z$, and the process repeats until $\tau=T$. The results are shown in Figure \ref{fig:SRTransient}.

The time domain algorithm is a perfectly valid one and is, in fact, more computationally efficient than either the ML algorithm or the integral Fourier (IF) algorithm (to be introduced in Section \ref{sec:if_MBEs}) for simulating our present narrow velocity distribution. Although this velocity extent is physically trivial, it spans a numerically non-trivial multiple ($21$) of the fundamental angular frequency differential $2\pi/T$ determined by the duration of the simulation. This experiment is thus a meaningful investigation of numerical accuracy; it is the goal of the IF algorithm of Section \ref{sec:if_MBEs} to enable, in future research, the efficient simulation of physically non-trivial velocity distributions which are otherwise intractable with the time domain representation of the MB equations.

\begin{center}
    \centering
    \begin{figure}
        \includegraphics[width=1.\columnwidth, trim=.5cm .75cm .5cm 2cm, clip]{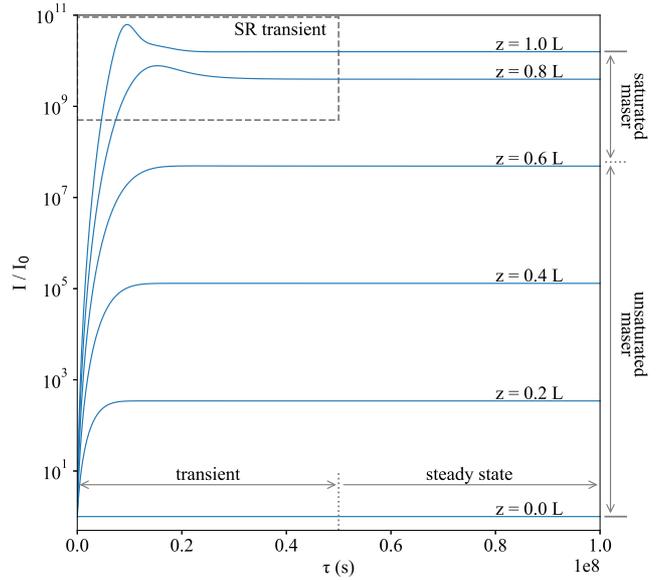}
        \caption{Intensity transients (normalized to the incident intensity $I_0$ at the start of the sample) at varying positions along a one-dimensional sample possessing a velocity distribution, inverted at $\tau=0$, and exhibiting SR at its end-fire ($z=L$). Computed from the time domain representation of the MB equations. Note the logarithmic scale for the intensity.}
        \label{fig:SRTransient}
    \end{figure}
\end{center}

\subsection{The Menegozzi \& Lamb algorithm in the quasi-steady state domain}\label{subsec:ss_MLr}

We first evaluate the ML algorithm with varying degrees of LMI fidelity against the quasi-steady state regime located on the right side of Figure \ref{fig:SRTransient}. We solve the ML equations \eqref{eq:FourierNpm}--\eqref{eq:zPropFour} via the procedure described in Section \ref{subsubsec:lmia} and with summations limited to $\bar{m} \in \left[-N_\text{int},+N_\text{int}\right]$ for varying values of the local mode interaction distance $N_\text{int}$.

The ML algorithm accurately converges to the steady state intensity profile; however, the LMI approximation fidelity requirements increase along the length of the sample as the system enters the saturated maser regions. Note that the saturated maser region can be identified as $z \gtrsim 0.6 \: L$, where the logarithm of the steady state intensity begins to deviate from constant-step increases when scanning vertically up the right side of Figure \ref{fig:SRTransient}. We summarise these findings by presenting the ML simulation of equations \eqref{eq:FourierNpm}--\eqref{eq:zPropFour} for our prototypical experiment in Figure \ref{fig:ML_SS}, computed with the LMI approximation at five degrees of fidelity, and in each case with spectral limiting (the range limitation on the mode index $m$ of each velocity channel) enforced to $m \in \left[-50, +50\right]$.

\begin{center}
    \centering
    \begin{figure}
        \includegraphics[width=1.\columnwidth, trim=.5cm 1.5cm 1cm 1.5cm, clip]{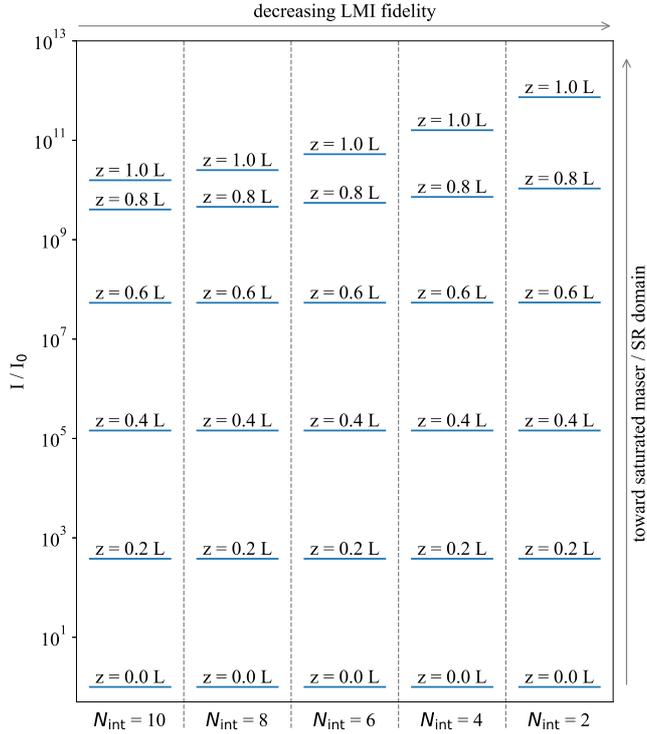}
        \caption{Quasi-steady state intensities (normalized to the incident intensity $I_0$ at the start of the sample) from the ML simulation of the system of Figure \ref{fig:SRTransient}, with LMI approximation interaction truncated to 10, 8, 6, 4, and 2 neighbouring modes. The fidelity requirements increase in the high field region $z>0.6L$, where the error in intensity becomes highly sensitive to reductions in $N_\text{int}$. For reference, the leftmost column ($N_\text{int}=10$) is effectively coincident with the true quasi-steady state values (cf. the right side of Figure \ref{fig:SRTransient}).}
        \label{fig:ML_SS}
    \end{figure}
\end{center}

The ML simulation converges to correct quasi-steady state intensities in the unsaturated maser domain across all degrees of LMI fidelity, but degrades with reduced LMI fidelity (moving to the right in Figure \ref{fig:ML_SS}) at greater $z$ positions (moving up in Figure \ref{fig:ML_SS}). This widened LMI mode coupling requirement at higher maser saturation suggests that an algorithm for simulating SR transients in the Fourier representation within regions of high field strength will also demand increased LMI approximation fidelity.

\subsection{The Menegozzi \& Lamb algorithm in the transient domain}\label{subsec:tr_MLr}

Although not naturally suited to modelling transient processes in its raw form of equations \eqref{eq:FourierNpm}--\eqref{eq:zPropFour}, the ML representation may be slightly revised to investigate the transient region on the left side of Figure \ref{fig:SRTransient}.

We begin this revision by noticing that the $m=0$ cases of equations \eqref{eq:FourierNpm} and \eqref{eq:FourierPpm} are, in fact, assertions of periodicity in the inversion and polarisation: when $m=0$, the right sides of equations \eqref{eq:FourierNpm} and \eqref{eq:FourierPpm} represent the zeroth modes of the Fourier series expansions of the right sides of equations \eqref{eq:MB_TD-1} and \eqref{eq:MB_TD-2}, which are the time derivatives of the inversion and polarisation. Generally speaking, the zeroth Fourier mode of a function is computed by integrating the function over the expansion domain $T$; thus, the $m=0$ equations are statements that the integral of the time derivatives of the inversion and polarisation must vanish; i.e., that they must be periodic in $T$.

This is a redundant statement to the assumption that the inversion and polarisation be represented, in the first place, by Fourier series expansions in the simulation duration $T$. We therefore drop the $m=0$ cases of equations \eqref{eq:FourierNpm} and \eqref{eq:FourierPpm}, and replace them with statements imposing our initial conditions upon the system in the ML representation; namely,
\begin{align}
    N_p\left(z,\tau=0\right) \equiv N_{p,0}\left(z\right) \quad
    &\Rightarrow \quad \sum_m \mathbb{N}_{p,m}\left(z\right) = N_{p,0}\left(z\right) \label{eq:ML_tr_ic_inv} \\
    \bar{\mathcal{P}}^\pm_p\left(z,\tau=0\right) \equiv \bar{\mathcal{P}}^\pm_{p,0}\left(z\right) \quad
    &\Rightarrow \quad \sum_m \bar{\mathbb{P}}^\pm_{p,m}\left(z\right) = \bar{\mathcal{P}}^\pm_{p,0}\left(z\right). \label{eq:ML_tr_ic_pol}
\end{align}
We refer to equations \eqref{eq:FourierNpm} and \eqref{eq:FourierPpm}, absolved of the $m=0$ case and augmented with the initial condition equations \eqref{eq:ML_tr_ic_inv} and \eqref{eq:ML_tr_ic_pol}, as the transient Menegozzi \& Lamb (TML) algorithm. The transients produced by the TML algorithm for our prototypical experiment are shown in Figure \ref{fig:TML_Transients}. Simulation in this case is executed with high LMI fidelity ($N_\text{int}=30$) and generous spectral limiting; i.e., $m \in \left[-50, +50\right]$.

The TML simulation demonstrates Gibbs ringing phenomena \citep{Bracewell1978} in positions advanced along the length of the sample, where SR processes yield substantial differences between the temporal initial and final configurations of the inversion and polarisation over the simulation duration $T$. Such ringing is present in any Fourier representation of a function which differs in value at the endpoints of the expansion interval \citep{Bracewell1978}; however, the overall form of the TML response is sensitive to this ringing and (problematically) renders this algorithm incapable of accurately modelling transient SR processes.

\begin{center}
    \centering
    \begin{figure}
        \includegraphics[width=1.\columnwidth, trim=.75cm 1.5cm 1.5cm 1.25cm, clip]{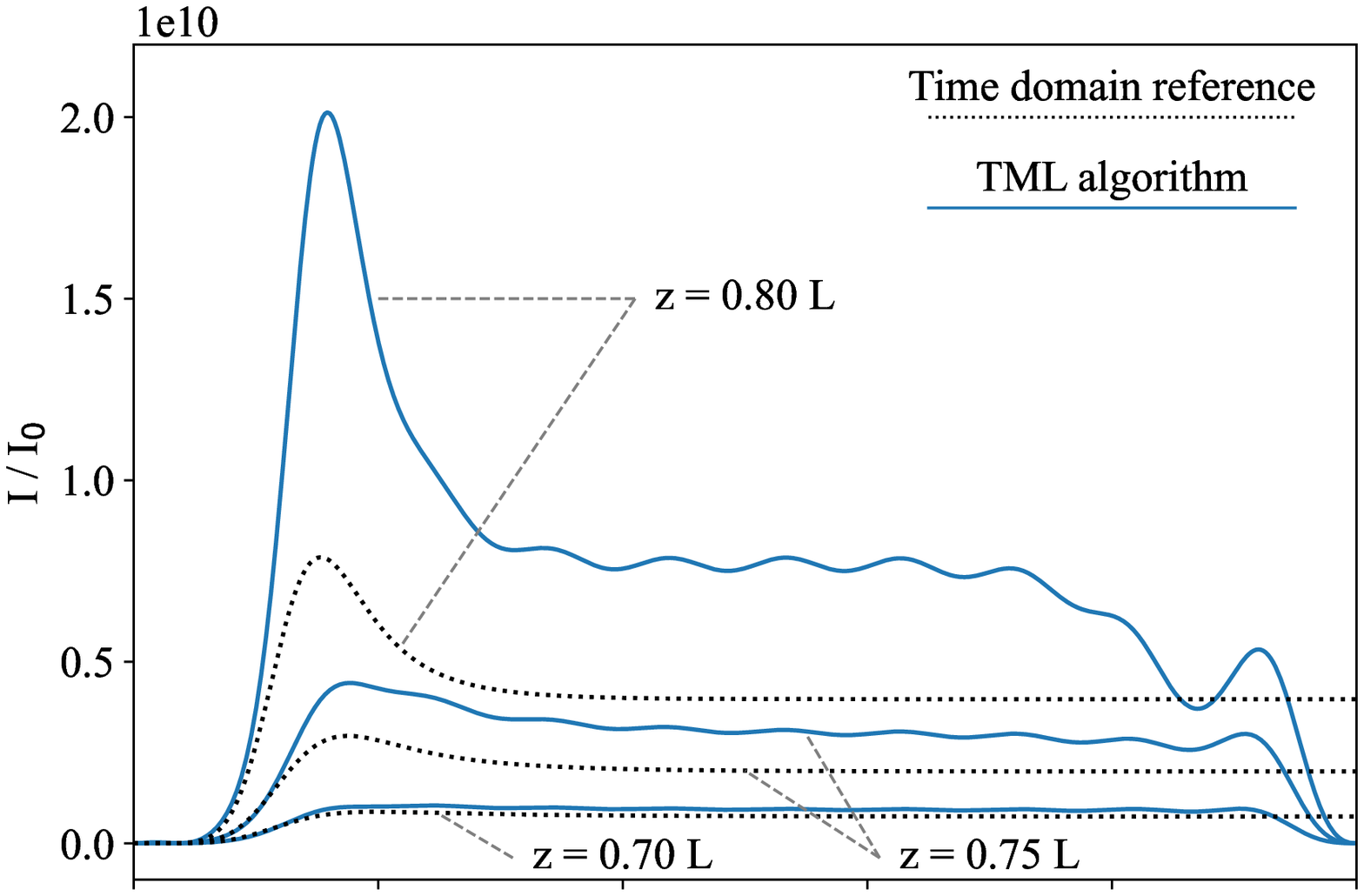}
        \includegraphics[width=1.\columnwidth, trim=.75cm 0cm 1.5cm 1.25cm, clip]{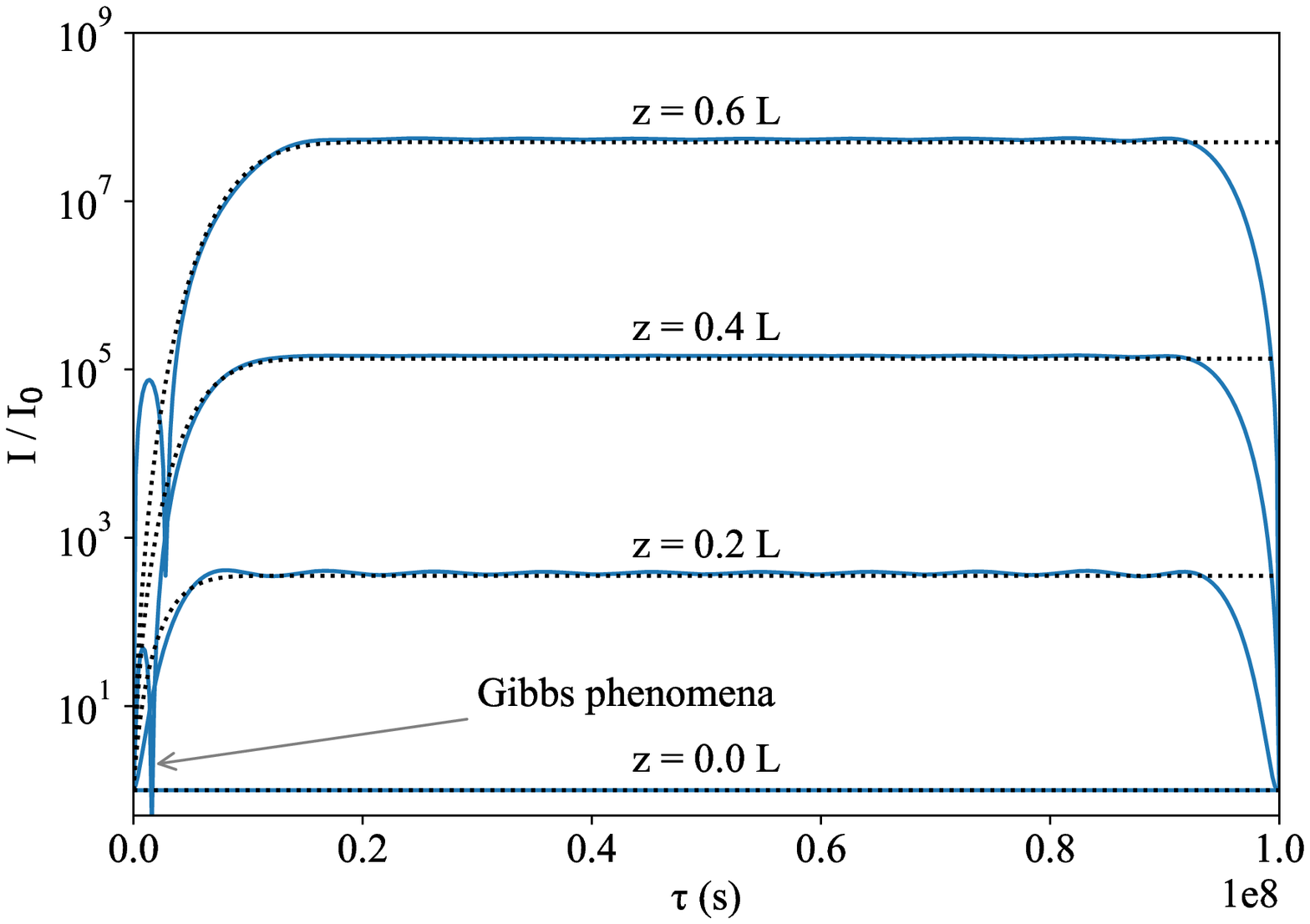}
        \caption{Intensity transients generated by a TML simulation of the experiment, with LMI approximation truncated to 30 modes. Top: transients shown for $0.70 L \leq z \leq 0.80 L$, linear vertical axis. Bottom: transients shown for $0.0L \leq z \leq 0.6 L$, logarithmic vertical axis. Gibbs ringing phenomena are visible in the logarithmic plots near the temporal boundaries.}
        \label{fig:TML_Transients}
    \end{figure}
\end{center}

The TML simulation appears reasonably capable of modelling the transient approach to the unsaturated maser steady states for $z \lesssim 0.6 \: L$, but begins to fail beyond $z \approx 0.6 \: L$ where the Gibbs ringing amplitude nears the magnitude of the imposed temporal initial conditions. For $z=0.8 \: L$ the peak amplitude near $\tau=1.6\times{10}^7 \text{ s}$ is incorrect by more than a factor of $2$ and the solution varies erratically in amplitude for all $\tau$ (note the logarithmic vertical axis of Figure \ref{fig:TML_Transients}). For $z \gtrsim 0.85$ the algorithm is completely unstable. Although the solutions at all $z$ correctly realise the temporal initial conditions enforced by equations \eqref{eq:ML_tr_ic_inv} and \eqref{eq:ML_tr_ic_pol}, asserting those conditions within a time interval demonstrating strong Gibbs ringing is a physically meaningless exercise. We turn now in Section \ref{sec:if_MBEs} to derive a new Fourier representation of the MB equations which yields meaningful transients immune to the temporal boundary Gibbs ringing phenomenon.

\section{The integral Fourier representation of the Maxwell-Bloch equations}\label{sec:if_MBEs}

In this section we derive a new Fourier representation of the MB equations that is manifestly distinct from the ML algorithm. This new representation is essentially the Fourier series of the integral form of the MB equations, and is therefore referred to as the \emph{integral Fourier} (IF) representation. Our IF representation naturally facilitates the enforcement of initial conditions, yields solutions converging to the optimal Fourier series representation of the correct time domain solution, contains a translation of the LMI approximation, and achieves $\mathcal{O}\left(N\right)$ complexity in the number of velocity channels $N$.

\subsection{The integral Fourier representation of a general first-order temporal propagation equation}\label{subsec:if_temp_prop}

We first derive the integral Fourier representation of a generic first-order temporal propagation problem, expressed as
\begin{equation}
    \frac{\mathrm{d}F}{\mathrm{d}t} = \mathcal{G} \left[ F\left(t\right), \: H\left(t\right) \right], \label{eq:Temporal_DiffEQ}
\end{equation}
where we seek to find the temporal propagation of $F$ from some specified initial conditions, given some generating expression $\mathcal{G}$ (which may potentially involve $F$) and (importantly) working exclusively within Fourier series expansions of the unknown quantities. Let $\mathcal{G}$ be expanded as
\begin{equation}
    \mathcal{G} \left(t\right) = \sum_{m} \mathbb{G}_{m} e^{i m d\!\omega t}, \label{eq:G_Frep}
\end{equation}
where $d\omega=2\pi/T$ for a simulation duration $T$, and where the $\mathbb{G}_{m}$ may potentially involve the coefficients $\mathbb{F}_{m}$ and $\mathbb{H}_{m}$ in the Fourier series expansions of the solution $F$ and the forcing function $H$,
\begin{align}
    F \left(t\right) &= \sum_{m} \mathbb{F}_{m}e^{i m d\!\omega t} \label{eq:F_Frep}\\
    H \left(t\right) &= \sum_{m} \mathbb{H}_{m}e^{i m d\!\omega t}. \label{eq:H_Frep}
\end{align}

The usual Fourier representation of equation (\ref{eq:Temporal_DiffEQ}) appropriate to steady state modelling is obtained by equating mode expansion coefficients upon insertion of equations \eqref{eq:G_Frep}--\eqref{eq:H_Frep} into equation (\ref{eq:Temporal_DiffEQ}). Motivated by our desire to introduce initial conditions into the Fourier representation, we instead first integrate both sides of equation (\ref{eq:Temporal_DiffEQ}); the left side introduces $F\left(t=0\right)$ as desired, while the right side we may analytically integrate in its Fourier representation. In equations, we have
\begin{align}
    F\left(t\right) - F\left(0\right) &= \int_{0}^{t} \mathrm{d} t' \sum_{m} \mathbb{G}_{m} e^{i m d\!\omega t'}\\
    &= \int_{0}^{t} \mathrm{d} t' \mathbb{G}_{0} + \sum_{m\neq0} \int_{0}^{t} \mathrm{d} t' \mathbb{G}_{m} e^{i m d\!\omega t'}\\
    &= \mathbb{G}_{0} t + \sum_{m\neq0} \frac{\mathbb{G}_{m}}{i m d\omega} \left(e^{i m d\!\omega t}-1\right).
\end{align}

Let us now express the function $t$ in a Fourier series as $t=\sum_{m}\mathbb{T}_{m}e^{imd\!\omega t}$ (we will explicitly compute $\mathbb{T}_{m}$ in a moment). We express $F$ on the left side in its Fourier series to obtain
\begin{equation}
    \begin{split}
        \sum_{m}\mathbb{F}_{m}e^{imd\!\omega t}-F\left(0\right) &= \mathbb{G}_{0}\sum_{m}\mathbb{T}_{m}e^{imd\!\omega t} \\
         &\quad + \sum_{m\neq0}\frac{\mathbb{G}_{m}}{imd\omega}\left(e^{imd\!\omega t}-1\right),
    \end{split}
\end{equation}
or
\begin{equation}
    \begin{split}
        &\mathbb{F}_{0} - F\left(0\right) + \sum_{m \neq 0}\mathbb{F}_{m}e^{imd\!\omega t} \\
        &\quad = \mathbb{G}_{0}\mathbb{T}_{0} - \sum_{m\neq0}\frac{\mathbb{G}_{m}}{imd\omega} + \sum_{m \neq 0}\left( \mathbb{G}_{0} \mathbb{T}_{m} +\frac{\mathbb{G}_{m}}{imd\omega} \right) e^{imd\!\omega t}.
    \end{split}
\end{equation}
Equating mode coefficients of equal frequencies generates the algebraic relations
\begin{align}
    \mathbb{F}_{0} &= F\left(0\right)+\mathbb{G}_{0}\mathbb{T}_{0}-\sum_{m\neq0}\frac{1}{i m d\omega}\mathbb{G}_{m}\label{eq:f0}\\
    \mathbb{F}_{m} &= \mathbb{G}_{0}\mathbb{T}_{m}+\frac{1}{i m d\omega}\mathbb{G}_{m}\text{ for }m\neq0.\label{eq:fm}
\end{align}
We compute $\mathbb{T}_{m}$ by the usual Fourier expansion coefficient calculation,
\begin{align}
    \mathbb{T}_{m} & =\frac{1}{T}\int_{0}^{T} t e^{-i m d\!\omega t} \mathrm{d} t\\
     &= \begin{cases}
        \frac{\pi}{d\omega} & m = 0\\
        \frac{i}{m d\omega} & m \neq 0.
    \end{cases}
\end{align}
Equations \eqref{eq:f0} and \eqref{eq:fm} now simplify to
\begin{align}
    \mathbb{F}_{0} &= F\left(0\right)+\sum_{m}\mathbb{T}_{m}\mathbb{G}_{m}\label{eq:f0_final} \\
    \mathbb{F}_{m} &= \mathbb{T}_{m}\left(\mathcal{\mathbb{G}}_{0}-\mathbb{G}_{m}\right) \text{ for }m \neq 0.\label{eq:fm_final}
\end{align} 

We refer to equations \eqref{eq:f0_final} and \eqref{eq:fm_final} as the integral Fourier (IF) representation of the first-order temporal propagation equation (\ref{eq:Temporal_DiffEQ}). Note that if the expression $\mathcal{G}$ of equation (\ref{eq:Temporal_DiffEQ}) contains $F$, then the $\mathbb{G}_{m}$ contain the $\mathbb{F}_{m}$ on the right sides of equations \eqref{eq:f0_final} and \eqref{eq:fm_final}, so that they must be solved as a linear system in the unknowns $\mathbb{F}_{m}$.

\subsection{The integral Fourier representation of the Maxwell-Bloch equations}\label{subsec:if_MBEs}

We seek now to cast the MB equations into the IF representation. Recognising the left sides of equations \eqref{eq:FourierNpm} and \eqref{eq:FourierPpm} as derivative operators acting on the inversion and polarisation, we can immediately infer from equations \eqref{eq:FourierNpm} and \eqref{eq:FourierPpm} that if
\begin{align}
    \frac{\partial N_{p}}{\partial\tau} &= \sum_{m} e^{i m d\!\omega \tau} \mathbb{G}_{m}^{(N)} \label{eq:Np_partial} \\
    \frac{\partial\bar{\mathcal{P}}_{p}^{+}}{\partial\tau} &= \sum_{m}e^{i m d\!\omega \tau} \mathbb{G}_m^{(P)},
\end{align}
then
\begin{align}
    \begin{split}
        \mathbb{G}_{m}^{(N)} &= \frac{i}{\hbar} \sum_{\bar{m}} \bigr( \mathcal{\bar{\mathbb{P}}}_{p,\bar{m}}^{+}\mathbb{E}_{\bar{m}-m+p}^{+} - \bar{\mathbb{P}}_{p,\bar{m}}^{-}\mathbb{E}_{\bar{m}+m+p}^{-} \bigr) \\
        &\quad -\frac{\mathbb{N}_{p,m}}{T_{1}}+\mathbb{L}_{m}^{(N)}
    \end{split} \\
    \mathbb{G}_{m}^{(P)} &= \frac{2id^{2}}{\hbar}\sum_{\bar{m}}\left(\mathbb{N}_{p,\bar{m}}\mathbb{E}_{m-\bar{m}+p}^{-}\right)-\frac{\mathcal{\bar{\mathbb{P}}}_{p,m}^{+}}{T_{2}}+\mathbb{L}_{m}^{(P)}. \label{eq:Gm_P}
\end{align}

Comparing equations \eqref{eq:Np_partial}--\eqref{eq:Gm_P} to the generic form of equation (\ref{eq:G_Frep}), we can immediately apply our IF representation formulae of equations \eqref{eq:f0_final} and \eqref{eq:fm_final} to obtain the IF representation of the MB equations,
\begin{align}
    \begin{split}
        \mathbb{N}_{p,0} &= N_{p}\left(0\right)+\sum_{m}\mathbb{T}_{m} \biggr[ \\
        &\quad \frac{i}{\hbar} \sum_{\bar{m}} \left(\bar{\mathbb{P}}_{p,\bar{m}}^{+}\mathbb{E}_{\bar{m}-m+p}^{+}-\bar{\mathbb{P}}_{p,\bar{m}}^{-}\mathbb{E}_{\bar{m}+m+p}^{-}\right) \\
        &\quad -\frac{\mathbb{N}_{p,m}}{T_{1}}+\mathbb{L}_{m}^{(N)}\biggr] \label{eq:MB_IF_1}
    \end{split}\\
    \begin{split}
        \mathbb{N}_{p,m\neq0} &= \mathbb{T}_{m} \biggr\{\frac{i}{\hbar} \sum_{\bar{m}} \bigr[\bar{\mathbb{P}}_{p,\bar{m}}^{+} \left(\mathbb{E}_{\bar{m}+p}^{+} - \mathbb{E}_{\bar{m}-m+p}^{+}\right) \\
        &\quad -\bar{\mathbb{P}}_{p,\bar{m}}^{-} \left(\mathbb{E}_{\bar{m}+p}^{-} - \mathbb{E}_{\bar{m}+m+p}^{-}\right) \bigr] \\
        &\quad +\frac{1}{T_{1}} \left(\mathbb{N}_{p,m} - \mathbb{N}_{p,0}\right) + \left(\mathbb{L}_{0}^{(N)} - \mathbb{L}_{m}^{(N)}\right)\biggr\} \label{eq:MB_IF_2}
    \end{split}\\
    \begin{split}
        \bar{\mathbb{P}}_{p,0}^{+} &= \bar{\mathcal{P}}_{p}^{+}\left(0\right) + \sum_{m}\mathbb{T}_{m} \biggr[\frac{2id^{2}}{\hbar}\sum_{\bar{m}}\left(\mathbb{N}_{p,\bar{m}}\mathbb{E}_{m-\bar{m}+p}^{-}\right) \\
        &\quad -\frac{\bar{\mathbb{P}}_{p,m}^{+}}{T_{2}} + \mathbb{L}_{m}^{(P)}\biggr] \label{eq:MB_IF_3}
    \end{split}\\
    \begin{split}
        \bar{\mathbb{P}}_{p,m\neq0}^{+} &= \mathbb{T}_{m} \biggr[\frac{2id^{2}}{\hbar} \sum_{\bar{m}}\mathbb{N}_{p,\bar{m}}\left(\mathbb{E}_{-\bar{m}+p}^{-}-\mathbb{E}_{m-\bar{m}+p}^{-}\right) \\
        &\quad +\frac{1}{T_{2}}\left(\bar{\mathbb{P}}_{p,m}^{+}-\bar{\mathbb{P}}_{p,0}^{+}\right)+\left(\mathbb{L}_{0}^{(P)}-\mathbb{L}_{m}^{(P)}\right)\biggr]. \label{eq:MB_IF_4}
    \end{split}
\end{align}
At a given $z$ position with known electric field modes $\mathbb{E}_{m}^{\pm}$, the above system of equations can be solved for the inversion and polarisation Fourier modes. Note that the electric field modes are propagated forward starting from $z=0$ via equation \eqref{eq:zPropFour}.

\subsubsection{The local mode interaction approximation and numerical considerations}\label{subsubsec:lmia_ifr}

The LMI approximation translates naturally to the IF representation under our Doppler shifted envelope factorisation of the polarisations. Equations \eqref{eq:MB_IF_1}--\eqref{eq:MB_IF_4} are formulated such that the LMI approximation is realised by restricting all occurrences of $\bar{m}$ to $\bar{m}\in\left[-N_{\text{int}},+N_{\text{int}}\right]$ for a desired mode interaction truncation distance $N_{\text{int}}$.

The spectral limiting of the $m$ index in each of $\mathbb{N}_{p,m}$ and $\bar{\mathbb{P}}_{p,m}$ achieves $\mathcal{O}\left(N\right)$ complexity in the number of velocity channels $N$; however, an additional efficiency is gained by exchanging summation orders in equations \eqref{eq:MB_IF_1} and \eqref{eq:MB_IF_3} and introducing the array
\begin{equation}
    \Xi_{a}^{\pm}=\sum_{m}\mathbb{T}_{m}\mathbb{E}_{a\mp m}^{\pm},\label{eq:Gamma_Def}
\end{equation}
such that equations \eqref{eq:MB_IF_1} and \eqref{eq:MB_IF_3} become
\begin{align}
    \begin{split}
        \mathbb{N}_{p,0} &= N_{p}\left(0\right) + \frac{i}{\hbar}\sum_{\bar{m}}\left(\Xi_{\bar{m}+p}^{+}\bar{\mathbb{P}}_{p,\bar{m}}^{+}-\Xi_{\bar{m}+p}^{-}\bar{\mathbb{P}}_{p,\bar{m}}^{-}\right) \\
        &\quad +\sum_{m}\mathbb{T}_{m}\left(\mathbb{L}_{m}^{(N)}-\frac{\mathbb{N}_{p,m}}{T_{1}}\right)\label{eq:IF_Xi_1}
    \end{split} \\
    \begin{split}
        \bar{\mathbb{P}}_{p,0}^{+} &= \bar{\mathcal{P}}_{p}^{+}\left(0\right)+\frac{2id^{2}}{\hbar}\sum_{\bar{m}}\Xi_{p-\bar{m}}^{-}\mathbb{N}_{p,\bar{m}} \\
        &\quad +\sum_{m}\mathbb{T}_{m}\left(\mathbb{L}_{m}^{(P)}-\frac{\bar{\mathbb{P}}_{p,m}^{+}}{T_{2}}\right).\label{eq:IF_Xi_2}
    \end{split}
\end{align}
At a given $z$ position, the $\Xi_{a}^{\pm}$ need only be calculated once, and may then be re-used to generate each linear system to be solved for each velocity channel of mode $p$. Note that because the inversion is real, its modes are related via $\mathbb{N}_{p,m}=\mathbb{N}_{p,-m}^{*}$, which reduces both the number of unknowns and the extents of inversion mode summations. After applying such simplifications, the real and imaginary representation of the IF system of equations is provided in Appendix \ref{app:if_mbes_re_n_im}.

It is helpful to define and clarify the ranges of all indices in equations (\ref{eq:IF_Xi_1}) and (\ref{eq:IF_Xi_2}), as well as in the definition of $\Xi_{a}^{\pm}$. We allow the velocity channel index $p$ to vary above and below the on-resonance $p=0$ central channel by the \emph{side channel} distance $N_\text{sch}$, so that $p\in\left[-N_{\text{sch}},+N_{\text{sch}}\right]$ for a total velocity channel count of $2 N_{\text{sch}}+1$. The mode index $m$ of the $p^\text{th}$ channel's inversion (or polarisation) expansion coefficients $\mathbb{N}_{p,m}$ ($\bar{\mathbb{P}}_{p,m}^{\pm}$) varies above and below the $m=0$ central mode by the \emph{side mode} distance $N_\text{sm}$ reflecting the degree of spectral limiting applied, so that $m\in\left[-N_{\text{sm}},+N_{\text{sm}}\right]$ for a total inversion (polarisation) mode count of $2 N_{\text{sm}}+1$. The electric field mode index $m$ of $\mathbb{E}_{m}^{\pm}$ is judiciously limited to $m\in\left[-\left(N_{\text{sch}}+N_{\text{sm}}\right),+\left(N_{\text{sch}}+N_{\text{sm}}\right)\right]$ for a total electric field mode count of $2\left(N_{\text{sch}}+N_{\text{sm}}\right)+1$. All occurrences of the index $\bar{m}$ vary over the \emph{interaction distance} $N_\text{int}$ such that $\bar{m}\in\left[-N_{\text{int}},+N_{\text{int}}\right]$. As a result of the above chosen ranges, the index $a$ of the $\Xi_{a}^{\pm}$ array extends over $a\in\left[-\left(N_{\text{sch}}+N_{\text{int}}\right),+\left(N_{\text{sch}}+N_{\text{int}}\right)\right]$.

\subsection{Simulation}\label{subsec:sim}

The result of simulating the same system of Figure \ref{fig:SRTransient} now with the IF algorithm is shown in Figure \ref{fig:IF_Transients}, where the inversion and polarisation are spectrally limited to $N_{\text{sm}} = 50$ and the LMI approximation is truncated to $N_{\text{int}} = 30$ neighbouring modes.

\begin{center}
    \centering
    \begin{figure}
        \includegraphics[width=1.\columnwidth, trim=.75cm 1.25cm 1.5cm 1cm, clip]{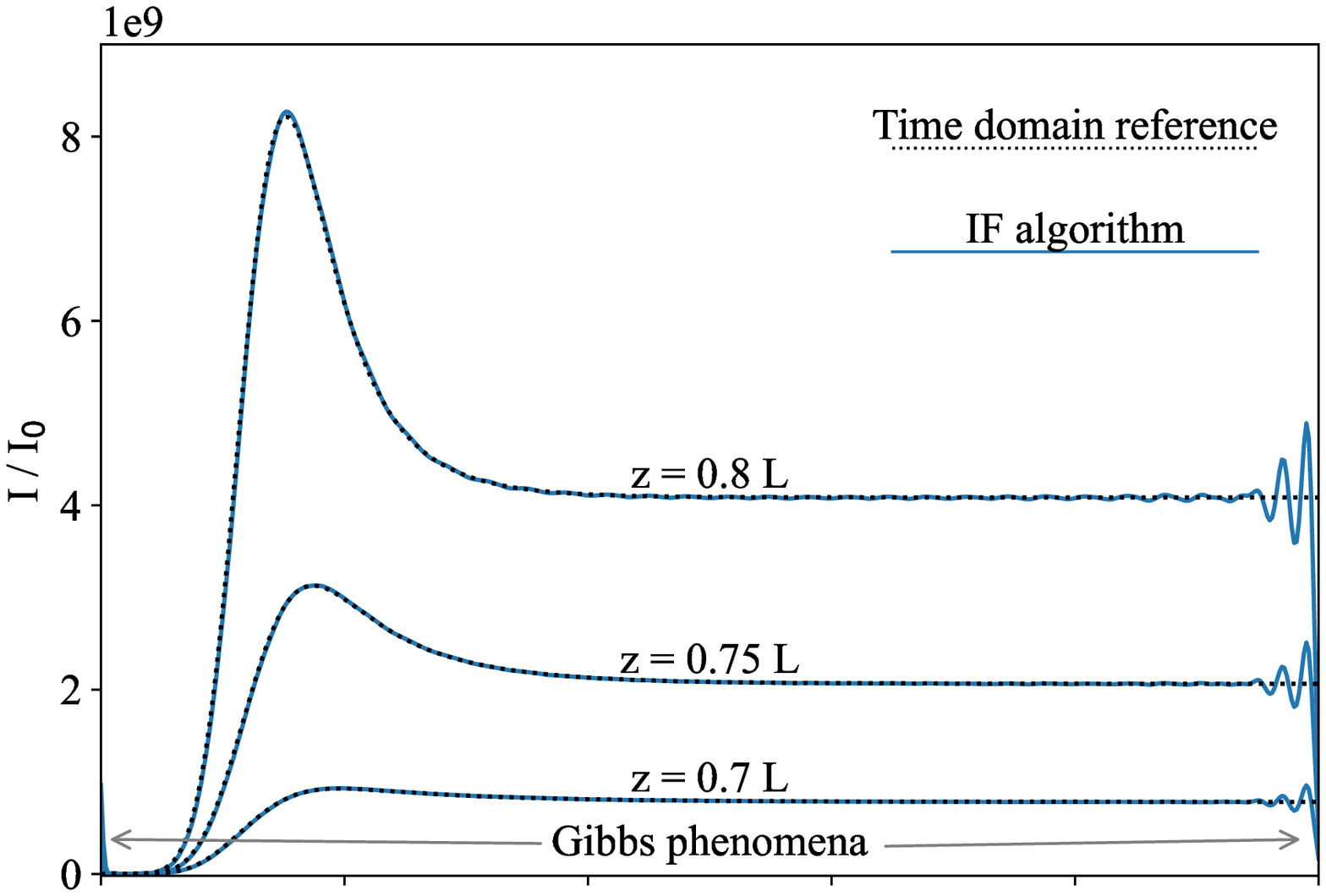}
        \includegraphics[width=1.\columnwidth, trim=.75cm 0cm 1.5cm 1.25cm, clip]{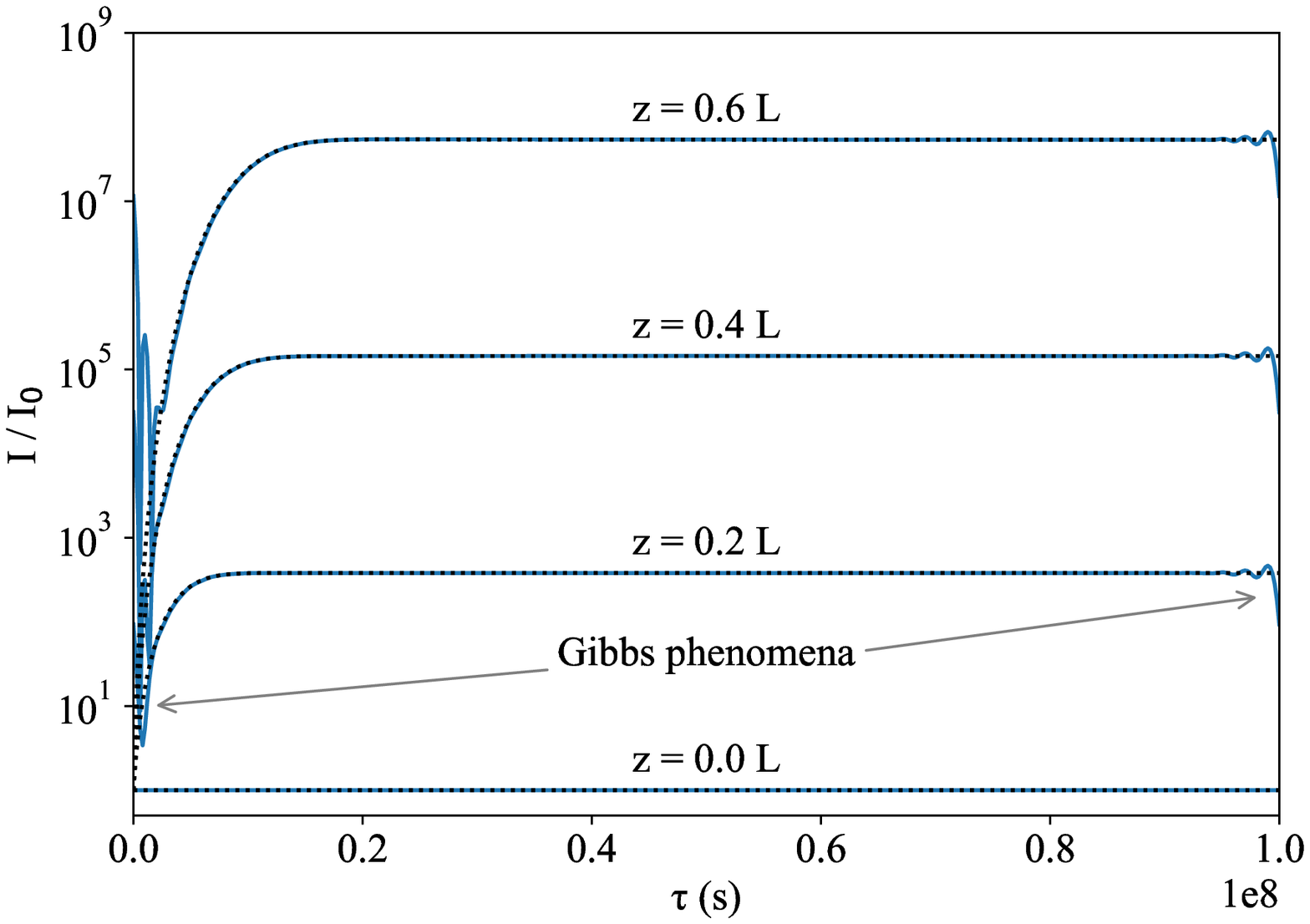}
        \caption{Intensity transients generated by an IF simulation of the experiment, with LMI approximation interaction truncated to 30 modes. Top: transients shown for $0.7 L \leq z \leq 0.8 L$, linear vertical axis. Bottom: transients shown for $0.0 L \leq z \leq 0.6 L$, logarithmic vertical axis. Reference time domain simulations are superimposed in dotted lines.}
        \label{fig:IF_Transients}
    \end{figure}
\end{center}

In contrast to the ML simulation of Figure \ref{fig:TML_Transients}, the IF simulation provides sustained accuracy into the SR transient domain (the upper left regions of Figures \ref{fig:IF_Transients} and \ref{fig:IF_Transients_Endfire}), where the important features of peak intensity magnitude and delay are properly recovered (compare to Figure \ref{fig:SRTransient}).

\begin{center}
    \centering
    \begin{figure}
        \includegraphics[width=1.\columnwidth, trim=1.25cm 0cm 1.5cm 1cm, clip]{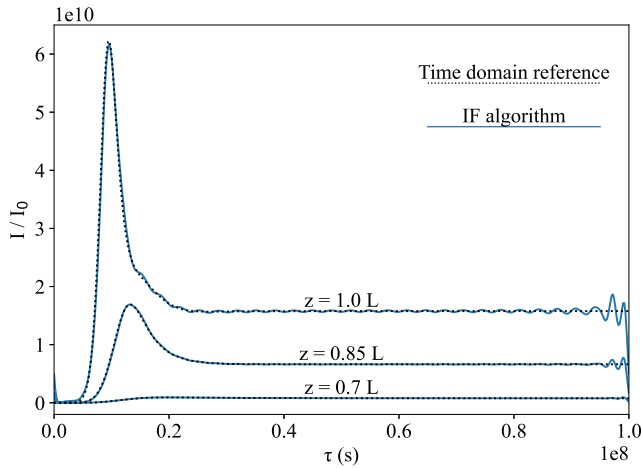}
        \caption{Intensity transients generated by an IF simulation of the experiment, with LMI approximation interaction truncated to 30 modes. Transients shown for $0.7 L \leq z \leq L$. Reference time domain simulations are superimposed in dotted lines.}
        \label{fig:IF_Transients_Endfire}
    \end{figure}
\end{center}

Despite the Gibbs ringing near the temporal boundaries of Figures \ref{fig:IF_Transients} and \ref{fig:IF_Transients_Endfire}, the total transient shape appears insensitive to such artifacts. The Gibbs phenomenon is also visible in the population inversion transients; we show in Figure \ref{fig:IF_inv_trans} the transients of the population inversion of the central velocity channel at various positions along the length of the sample. The accuracy of the IF algorithm is again verified in Figure \ref{fig:IF_inv_trans} where, importantly, the total response is unaffected by the aggressive Gibbs ringing at the boundaries. For completeness, we show in Figure \ref{fig:IF_pol_trans} the imaginary part of the polarisation of the central velocity channel at various positions along the length of the sample (in this particular simulation the real part of the central velocity channel's polarisation is negligible).

\begin{center}
    \centering
    \begin{figure}
        \includegraphics[width=1.\columnwidth, trim=0.75cm 0cm 1.5cm 1cm, clip]{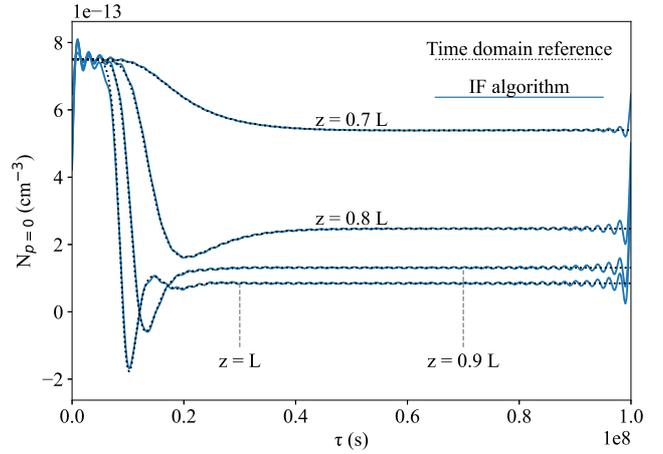}
        \caption{Population inversion transients generated by an IF simulation of the experiment, with LMI approximation interaction truncated to 30 modes. Reference time domain simulations are superimposed in dotted lines. Note that the molecular density used for the $\Delta v$ of this simulation corresponds to an inverted molecular density on the order of $0.1 \text{ cm}^{-3}$ for a realistic velocity distribution of $\Delta v \approx 1 \text{ km/s}$.}
        \label{fig:IF_inv_trans}
    \end{figure}
\end{center}

\begin{center}
    \centering
    \begin{figure}
        \includegraphics[width=1.\columnwidth, trim=.5cm 0cm 1.5cm 1cm, clip]{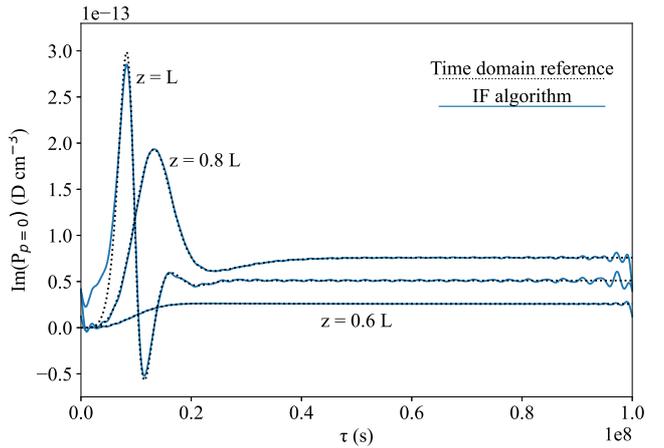}
        \caption{Imaginary part of polarisation transients generated by an IF simulation of the experiment, with LMI approximation interaction truncated to 30 modes. Reference time domain simulations are superimposed in dotted lines.}
        \label{fig:IF_pol_trans}
    \end{figure}
\end{center}

\subsection{Local mode interaction fidelity requirements}

We turn now to quantify the effect of reducing the LMI fidelity upon the accuracy of the SR transients generated. In Figure \ref{fig:IF_LMI_exp} we plot the intensity transients for varying LMI truncation extents at two positions along the length of the sample. At $z=0.6 L$ the system does not yet demonstrate SR; i.e., there is no loss of population inversion until $z \approx 0.7 L$, as can be seen in Figure \ref{fig:IF_inv_trans}. Conversely, at $z=L$ the system generates a strong SR pulse with significant and fast variations in the population inversion, as well as a peak SR intensity greatly exceeding the steady state value.

\begin{center}
    \centering
    \begin{figure}
        \includegraphics[width=1.\columnwidth, trim=1.25cm 1.30cm 1.75cm 1cm, clip]{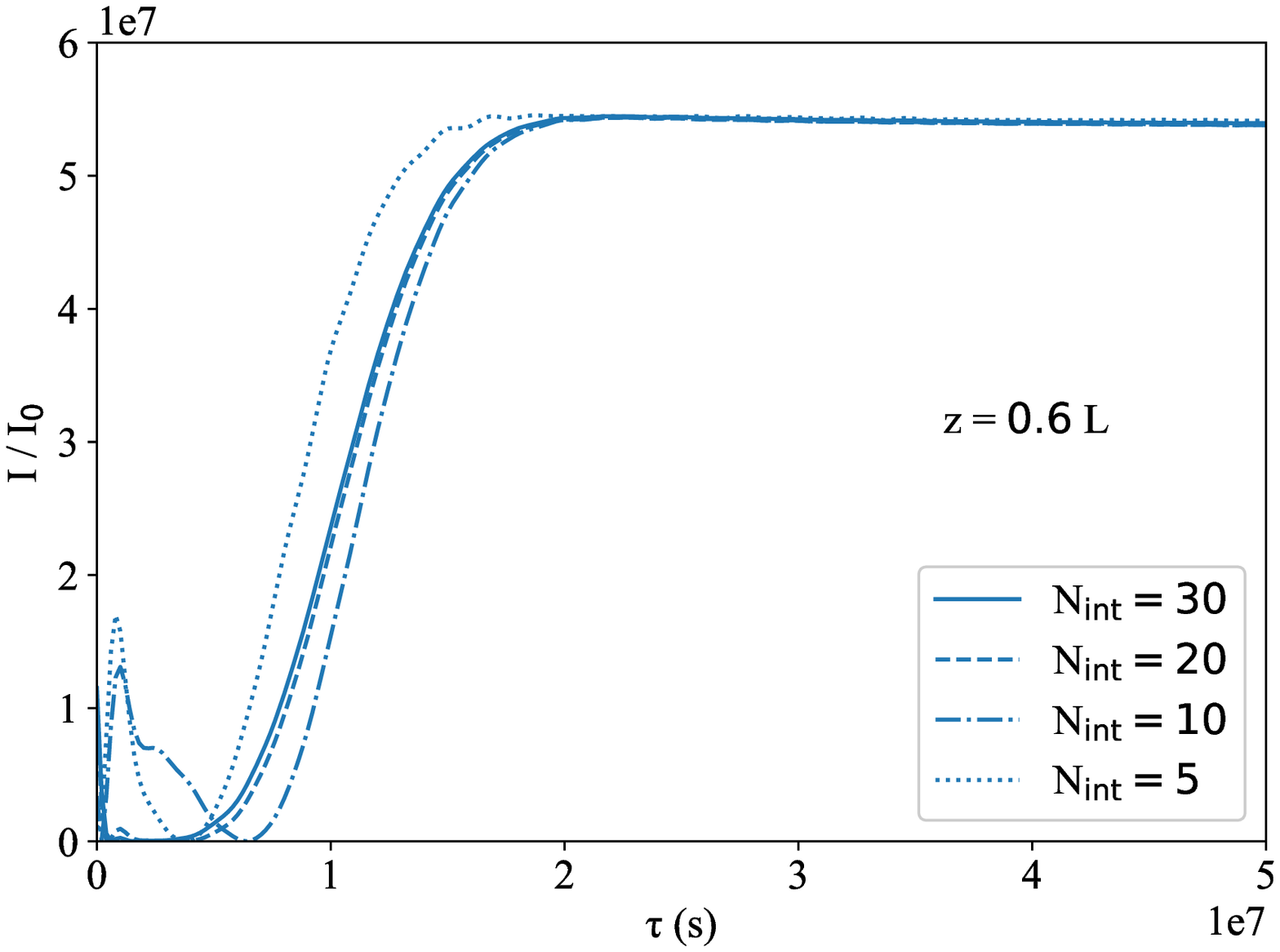}
        \includegraphics[width=1.\columnwidth, trim=1.25cm 0cm 1.75cm 1cm, clip]{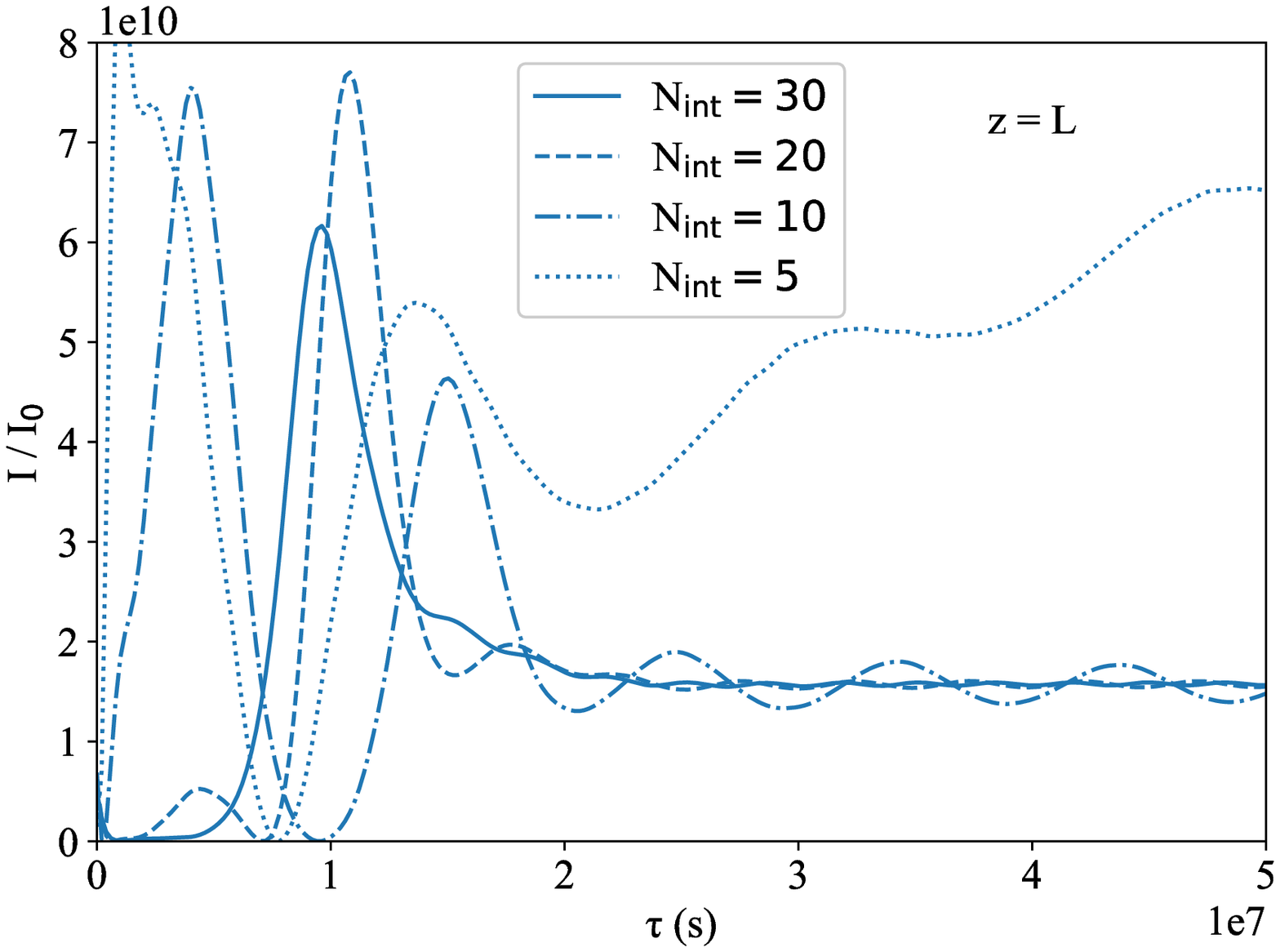}
        \caption{IF simulation intensity transients at $z = 0.6 L$ (top) and $z = L$ (bottom) for varying degrees of LMI approximation fidelity. The $N_\text{int}=30$ plots (solid lines) may serve as reference true transients, being effectively coincident with the (correct) transients generated by a time domain simulation. Significant deviations begin to emerge below $N_\text{int} \approx 5$ at $z=0.6 L$ (low field strength), and below $N_\text{int} \approx 20$ at $z=L$ (high field strength).}
        \label{fig:IF_LMI_exp}
    \end{figure}
\end{center}

We observe that the peak intensity at $z = 0.6 L$ remains very accurate (it tracks well with the $N_\text{int}=30$ case) down to an LMI truncation extent as low as $N_\text{int} \approx 10$, and reasonably accurate down to $N_\text{int} \approx 5$. Conversely, the SR transient at $z = L$ acquires a moderate error already at $N_\text{int}=20$, a significant error at $N_\text{int}=10$, and becomes completely unstable at $N_\text{int}=5$. These observations suggest that the LMI fidelity requirements increase as the system enters the transient SR domain.

Let us compare the bandwidth of the SR pulse in the bottom plot of Figure \ref{fig:IF_LMI_exp} to the LMI truncation extent. The pulse has a duration on the order of $\sim\! T / 20$ (note that $T=1\times{10}^8 \text{ s}$, despite the reduced plot viewing widths), or a bandwidth of $\Delta \omega \approx 20 \left(2 \pi / T\right)$. The natural angular frequencies of adjacent velocity channels are separated by an amount $2 \pi / T$, according to our prescription of a channel separation $dv = \left(c/\omega_0\right) 2\pi / T$. The bandwidth of the SR pulse thus covers the bandwidth of approximately $20$ velocity channels, which is the value of $N_\text{int}$ at which the approximated transient begins to depart from the true transient. This suggests that the LMI velocity channel truncation extent, interpreted according to the equivalent Doppler shift across the interacting channels, must exceed the bandwidth of the transient generated.

\section{Discussion}\label{sec:Disc_n_future}

\subsection{Conclusions}\label{subsec:Conclusions}

The ML algorithm was originally developed by \citet{Menegozzi1978} (and extended by \citealt{Trung2009a} and \citealt{Trung2009b}) to investigate noise propagation and formation of coherence in a quasi-steady state maser process. Motivated by the advantageous $\mathcal{O}\left(N\right)$ complexity (in the velocity channel count $N$) achieved by the ML algorithm over the $\mathcal{O}\left(N^2\right)$ complexity of a time domain simulation of the MB equations with a velocity distribution, we have investigated the application of the ML algorithm to the modelling of SR transient processes.

The ML algorithm accurately describes the quasi-steady state maser regime. After a minor revision to its algebraic mode relations (a replacement of the $m=0$ cases with initial condition assertions), it also describes weak field transients (see $z \lesssim 0.6 L$ in Figure \ref{fig:TML_Transients}); however, we have demonstrated it unacceptably sensitive to Gibbs phenomena in the case of strong field SR transients ($z \gtrsim 0.7 L$ in Figure \ref{fig:TML_Transients}). The transient ML algorithm does not accurately converge to the true transient if the amplitude of the Gibbs ringing exceeds the magnitude of the initial conditions asserted, and it is therefore unable to describe SR transient processes where the loss in population inversion caused by the SR cascade causes the temporal initial and final configurations to differ substantially.

We have developed a manifestly unique Fourier representation of the MB equations which lends itself naturally to the assertion of initial conditions and which accurately models all SR transient processes described by the MB equations. The IF algorithm is robustly insensitive to Gibbs phenomena, yielding total transients which accurately replicate key SR features such as peak intensity delay time and peak intensity amplitude. Most importantly, the spectral limiting and LMI approximations of the ML algorithm translate naturally into the IF algorithm, so that the latter is also $\mathcal{O}\left(N\right)$ complex in the number of velocity channels $N$.

We have observed that the fidelity requirements of the LMI approximation made in either the ML representation or the IF representation of the MB equations increase as a system approaches high field strength regimes. We suggest that the fidelity of the LMI approximation used when simulating SR transients in the IF representation be such that the natural frequency extent of the spread in velocity channel interactions exceeds the bandwidth of the transient response.

\subsection{Limitations of the Integral Fourier Method}\label{subsec:limitations}

In a footnote to Section \ref{subsec:sp_em_n_init_bloch}, it was emphasised that although the initial Bloch tipping angle is prescribed by the number of interacting molecules $N_\text{mol int}$, such number is not known \emph{a priori} when simulating across a wide velocity distribution. In fact, $N_\text{mol int}$ could more accurately be estimated from simulations enabled by this paper, which would provide an estimate of the local number of coherent interacting neighbouring molecules within a subset of the full global velocity distribution. Alternatively, the purely quantum mechanical arguments of \citet{Gross1982} which inform the initial Bloch angle prescription could be re-examined in the context of a velocity distribution, with the objective of obtaining a revised expression for $N_\text{mol int}$. In most astrophysical situations, however, other radiative processes are of sufficient intensity to render the initial tipping angle irrelevant. In \citet{Rajabi2020}, for example, the background radiation of the interstellar medium is shown to dominate over any initial radiation field resulting from the non-zero initial tipping angle.

On a numerical efficiency note, it should be pointed out that although the IF algorithm achieves $\mathcal{O}\left(N\right)$ complexity in the number of velocity channels $N$ simulated (compared to $\mathcal{O}\left(N^2\right)$ for the time domain method), it does not overtake the time domain method until approximately $N>100$. The simplicity of the time domain method, owing in part to its lack of matrix inversion operations, therefore makes it the superior choice of algorithm for simulating nearly coherent velocity distributions.

\subsection{Future Work}\label{subsec:future_work}

This paper is a proof of operation of the IF algorithm, and all simulations were performed over a velocity extent demanding a numerical simulation duration as brief as $\sim\! 40$ seconds when solving in the time domain on a modern (2020 quad core) CPU over a $500\times500$-point $\left(n_\tau \times n_z\right)$ grid. These velocity extents are numerically non-trivial and serve as valid demonstrations of proper convergence to the reference time domain solution, but at present remain physically trivial ($\sim\!{10}^{-6} \text{ cm/s}$ for the samples simulated in Section \ref{sec:if_MBEs}). This paper enables future research to apply the $\mathcal{O}\left(N\right)$ complex IF algorithm to physically realistic velocity distributions for the study of SR processes in astrophysical gases.

The objective of future simulations will be to characterise the degree of coherence expected of transient astrophysical SR events generated from wide velocity distributions of molecules. It is the hope that the coherence characteristics predicted by these simulations will constitute a metric by which to ascertain the presence of transient SR events within observational data.

Simulations of SR processes across wide velocity distributions will also provide realistic corrections to physical requirements for a system to demonstrate SR, as well as to characteristic features of SR transients. Such parameters include the critical column density threshold required to initiate an SR transient event, the time delay $\tau_\text{D}$ to peak intensity of the SR pulse, and the characteristic time scale $T_\text{R}$ for the dissipation of energy from the system.

Additionally, the IF simulation enables investigations of noise propagation along the length of a sample during a transient SR process. In the original work of \citet{Menegozzi1978}, an incident electric field possessing a broad and decoherent spectrum of Gaussian white noise was set incident upon the $z=0$ face of a one-dimensional maser sample, and the coherence of the emerging radiation at the end-fire $z=L$ was evaluated. Similarly, the IF simulation enables future research to investigate the relationship between the statistics of incident radiation noise at $z=0$ and the coherence of emerging radiation at $z=L$ when the relevant collective emission process occurring within the sample is that of a transient SR event.

Finally, a possible generalisation of the IF algorithm introduced briefly near the end of Section \ref{subsubsec:lmia} warrants elaboration. The simulations of this paper were executed with a fixed LMI approximation fidelity for all $z$; however, the velocity channel interaction distance $N_\textrm{int}$ could, in theory, be made to vary as a function of $z$ (or more precisely, as a function of the local degree of saturation). The present code fixes a global $N_\textrm{int}$ according to the highest degree of saturation occurring at $z=L$. Conversely, for $z$ moderately less than $L$, lower degrees of saturation reduce the LMI approximation fidelity requirements. A reduction in $N_\textrm{int}$ at such positions would reduce the numerical complexity of solving the coupled system of equations \eqref{eq:MB_IF_1}--\eqref{eq:MB_IF_4} within the majority of the sample, and could feasibly improve computation speed by an order of magnitude.

\section*{Acknowledgements}
C.M.W. is supported by the Natural Sciences and Engineering Research Council of Canada (NSERC) through the doctoral postgraduate scholarship (PGS D). B.L. acknowledges support from the Swedish Research Council (VR) through grant No. 2014-05713. F.R.’s research at Perimeter Institute is supported in part by the Government of Canada through the Department of Innovation, Science and Economic Development Canada and by the Province of Ontario through the Ministry of Economic Development, Job Creation and Trade. M.H.'s research is funded through the Natural Sciences and Engineering Research Council of Canada Discovery Grant RGPIN-2016-04460 and the Western Strategic Support for NSERC Success Accelerator program.

\section*{Data Availability Statement}
The data pipeline is made available at: \url{https://github.com/cwyenberg/MandL-to-Superradiance} and maintained by C.M.W. The figures in this paper were prepared using the {\tt matplotlib} package \citep{Hunter2007}.




\bibliographystyle{mnras}
\bibliography{scibib}

\begin{thebibliography}{}
\makeatletter
\relax
\def\mn@urlcharsother{\let\do\@makeother \do\$\do\&\do\#\do\^\do\_\do\%\do\~}
\def\mn@doi{\begingroup\mn@urlcharsother \@ifnextchar [ {\mn@doi@}
  {\mn@doi@[]}}
\def\mn@doi@[#1]#2{\def\@tempa{#1}\ifx\@tempa\@empty \href
  {http://dx.doi.org/#2} {doi:#2}\else \href {http://dx.doi.org/#2} {#1}\fi
  \endgroup}
\def\mn@eprint#1#2{\mn@eprint@#1:#2::\@nil}
\def\mn@eprint@arXiv#1{\href {http://arxiv.org/abs/#1} {{\tt arXiv:#1}}}
\def\mn@eprint@dblp#1{\href {http://dblp.uni-trier.de/rec/bibtex/#1.xml}
  {dblp:#1}}
\def\mn@eprint@#1:#2:#3:#4\@nil{\def\@tempa {#1}\def\@tempb {#2}\def\@tempc
  {#3}\ifx \@tempc \@empty \let \@tempc \@tempb \let \@tempb \@tempa \fi \ifx
  \@tempb \@empty \def\@tempb {arXiv}\fi \@ifundefined
  {mn@eprint@\@tempb}{\@tempb:\@tempc}{\expandafter \expandafter \csname
  mn@eprint@\@tempb\endcsname \expandafter{\@tempc}}}

\bibitem[\protect\citeauthoryear{Andreev}{Andreev}{1990}]{Andreev1990}
Andreev A.~V.,  1990, Soviet Physics Uspekhi, 33, 997

\bibitem[\protect\citeauthoryear{{Arecchi} \& {Courtens}}{{Arecchi} \&
  {Courtens}}{1970}]{Arecchi1970}
{Arecchi} F.~T.,  {Courtens} E.,  1970, \mn@doi [\pra]
  {10.1103/PhysRevA.2.1730}, \href
  {http://adsabs.harvard.edu/abs/1970PhRvA...2.1730A} {2, 1730}

\bibitem[\protect\citeauthoryear{Benedict et~al.}{Benedict
  et~al.}{1996}]{Benedict1996}
Benedict M.~G.,  et~al., 1996, Super-radiance: Multiatomic Coherent Emission.
IOP Publishing Ltd

\bibitem[\protect\citeauthoryear{Bracewell}{Bracewell}{1978}]{Bracewell1978}
Bracewell R.,  1978, The Fourier Transform and its Applications, second edn.
McGraw-Hill Kogakusha, Ltd., Tokyo

\bibitem[\protect\citeauthoryear{{Caratti o Garatti} et~al.,}{{Caratti o
  Garatti} et~al.}{2017}]{Caratti2017}
{Caratti o Garatti} A.,  et~al., 2017, \mn@doi [Nature Physics]
  {10.1038/nphys3942}, \href
  {http://adsabs.harvard.edu/abs/2017NatPh..13..276C} {13, 276}

\bibitem[\protect\citeauthoryear{Dicke}{Dicke}{1954}]{Dicke1954}
Dicke R.~H.,  1954, \mn@doi [Phys. Rev.] {10.1103/PhysRev.93.99}, \href
  {http://adsabs.harvard.edu/abs/1954PhRv...93...99D} {93, 99}

\bibitem[\protect\citeauthoryear{Dinh-V-Trung}{Dinh-V-Trung}{2009a}]{Trung2009a}
Dinh-V-Trung 2009a, \mn@doi [MNRAS] {10.1111/j.1365-2966.2009.14901.x}, 396,
  2319

\bibitem[\protect\citeauthoryear{Dinh-V-Trung}{Dinh-V-Trung}{2009b}]{Trung2009b}
Dinh-V-Trung 2009b, \mn@doi [MNRAS] {10.1111/j.1365-2966.2009.15369.x}, 399,
  1495

\bibitem[\protect\citeauthoryear{{Elitzur}}{{Elitzur}}{1992}]{Elitzur1992}
{Elitzur} M.,  1992, {Astronomical Masers}.
Springer, Dordrecht, \mn@doi{10.1007/978-94-011-2394-5}

\bibitem[\protect\citeauthoryear{Feld \& MacGillivray}{Feld \&
  MacGillivray}{1980}]{Feld1980}
Feld M.,  MacGillivray J.,  1980, in , Coherent Nonlinear Optics.
Springer, pp 7--57

\bibitem[\protect\citeauthoryear{{Field} \& {Gray}}{{Field} \&
  {Gray}}{1988}]{Field1988}
{Field} D.,  {Gray} M.~D.,  1988, \mn@doi [\mnras] {10.1093/mnras/234.2.353},
  \href {https://ui.adsabs.harvard.edu/abs/1988MNRAS.234..353F} {234, 353}

\bibitem[\protect\citeauthoryear{{Field} \& {Richardson}}{{Field} \&
  {Richardson}}{1984}]{Field1984}
{Field} D.,  {Richardson} I.~M.,  1984, \mn@doi [\mnras]
  {10.1093/mnras/211.4.799}, \href
  {https://ui.adsabs.harvard.edu/abs/1984MNRAS.211..799F} {211, 799}

\bibitem[\protect\citeauthoryear{Goldreich \& Keeley}{Goldreich \&
  Keeley}{1972}]{Goldreich1972}
Goldreich P.,  Keeley D.~A.,  1972, \mn@doi [The Astrophysical Journal]
  {10.1086/151514}, 174, 517

\bibitem[\protect\citeauthoryear{{Goldreich} \& {Kwan}}{{Goldreich} \&
  {Kwan}}{1974}]{Goldreich1974}
{Goldreich} P.,  {Kwan} J.,  1974, \mn@doi [\apj] {10.1086/152843}, \href
  {http://adsabs.harvard.edu/abs/1974ApJ...190...27G} {190, 27}

\bibitem[\protect\citeauthoryear{{Gray}}{{Gray}}{2012}]{Gray2012}
{Gray} M.,  2012, {Maser Sources in Astrophysics}.
Cambridge University Press, Cambridge, UK

\bibitem[\protect\citeauthoryear{Gross \& Haroche}{Gross \&
  Haroche}{1982}]{Gross1982}
Gross M.,  Haroche S.,  1982, \physrep, 93, 301

\bibitem[\protect\citeauthoryear{Houde, Mathews  \& Rajabi}{Houde
  et~al.}{2018}]{Houde2018a}
Houde M.,  Mathews A.,   Rajabi F.,  2018, \mn@doi [\mnras]
  {10.1093/mnras/stx3205}, \href
  {http://adsabs.harvard.edu/abs/2018MNRAS.475..514H} {475, 514}

\bibitem[\protect\citeauthoryear{{Houde}, {Rajabi}, {Gaensler}, {Mathews}  \&
  {Tranchant}}{{Houde} et~al.}{2019}]{Houde2019}
{Houde} M.,  {Rajabi} F.,  {Gaensler} B.~M.,  {Mathews} A.,   {Tranchant} V.,
  2019, \mn@doi [\mnras] {10.1093/mnras/sty3046}, \href
  {http://adsabs.harvard.edu/abs/2019MNRAS.482.5492H} {482, 5492}

\bibitem[\protect\citeauthoryear{Hunter}{Hunter}{2007}]{Hunter2007}
Hunter J.~D.,  2007, \mn@doi [Computing in Science \& Engineering]
  {10.1109/MCSE.2007.55}, 9, 90

\bibitem[\protect\citeauthoryear{Litvak}{Litvak}{1970}]{Litvak1970}
Litvak M.~M.,  1970, \mn@doi [Phys. Rev. A] {10.1103/PhysRevA.2.2107}, 2, 2107

\bibitem[\protect\citeauthoryear{{MacGillivray} \& {Feld}}{{MacGillivray} \&
  {Feld}}{1976}]{MacGillivray1976}
{MacGillivray} J.~C.,  {Feld} M.~S.,  1976, \mn@doi [\pra]
  {10.1103/PhysRevA.14.1169}, \href
  {https://ui.adsabs.harvard.edu/abs/1976PhRvA..14.1169M} {14, 1169}

\bibitem[\protect\citeauthoryear{Mathews}{Mathews}{2017}]{Mathews2017}
Mathews A.,  2017, The Role of Superradiance in Cosmic Fast Radio Bursts,
  Honours thesis, The University of Western Ontario

\bibitem[\protect\citeauthoryear{Menegozzi \& Lamb}{Menegozzi \&
  Lamb}{1978}]{Menegozzi1978}
Menegozzi L.~N.,  Lamb W.~E.,  1978, \mn@doi [Phys. Rev. A]
  {10.1103/PhysRevA.17.701}, 17, 701

\bibitem[\protect\citeauthoryear{Polder, Schuurmans  \& Vrehen}{Polder
  et~al.}{1979}]{Polder1979}
Polder D.,  Schuurmans M.,   Vrehen Q.,  1979, Physical Review A, 19, 1192

\bibitem[\protect\citeauthoryear{Rajabi \& Houde}{Rajabi \&
  Houde}{2016a}]{Rajabi2016A}
Rajabi F.,  Houde M.,  2016a, \apj, 826, 216

\bibitem[\protect\citeauthoryear{Rajabi \& Houde}{Rajabi \&
  Houde}{2016b}]{Rajabi2016B}
Rajabi F.,  Houde M.,  2016b, \apj, 828, 57

\bibitem[\protect\citeauthoryear{{Rajabi} \& {Houde}}{{Rajabi} \&
  {Houde}}{2017}]{Rajabi2017}
{Rajabi} F.,  {Houde} M.,  2017, \mn@doi [Sci. Adv.] {10.1126/sciadv.1601858},
  \href {http://adsabs.harvard.edu/abs/2017SciA....3E1858R} {3, e1601858}

\bibitem[\protect\citeauthoryear{{Rajabi} \& {Houde}}{{Rajabi} \&
  {Houde}}{2020}]{Rajabi2020}
{Rajabi} F.,  {Houde} M.,  2020, \mn@doi [\mnras] {10.1093/mnras/staa1067},
  \href {https://ui.adsabs.harvard.edu/abs/2020MNRAS.494.5194R} {494, 5194}

\bibitem[\protect\citeauthoryear{{Rajabi}, {Houde}, {Bartkiewicz}, {Olech},
  {Szymczak}  \& {Wolak}}{{Rajabi} et~al.}{2019}]{Rajabi2019}
{Rajabi} F.,  {Houde} M.,  {Bartkiewicz} A.,  {Olech} M.,  {Szymczak} M.,
  {Wolak} P.,  2019, \mn@doi [\mnras] {10.1093/mnras/stz074}, \href
  {https://ui.adsabs.harvard.edu/abs/2019MNRAS.484.1590R} {484, 1590}

\bibitem[\protect\citeauthoryear{{Rajabi}, {Chamma}, {Wyenberg}, {Mathews}  \&
  {Houde}}{{Rajabi} et~al.}{2020}]{Rajabi2020b}
{Rajabi} F.,  {Chamma} M.~A.,  {Wyenberg} C.~M.,  {Mathews} A.,   {Houde} M.,
  2020, \mn@doi [\mnras] {10.1093/mnras/staa2723}, \href
  {https://ui.adsabs.harvard.edu/abs/2020MNRAS.498.4936R} {498, 4936}

\bibitem[\protect\citeauthoryear{Sargent, Scully  \& Lamb}{Sargent
  et~al.}{1974}]{Sargent1974}
Sargent M.~I.,  Scully M.~O.,   Lamb W. E.~j.,  1974, Laser physics.
Addison-Wesley, 1974

\bibitem[\protect\citeauthoryear{{Sobolev}, {Cragg}  \& {Godfrey}}{{Sobolev}
  et~al.}{1997}]{Sobolev1997}
{Sobolev} A.~M.,  {Cragg} D.~M.,   {Godfrey} P.~D.,  1997, \aap, \href
  {http://adsabs.harvard.edu/abs/1997A%26A...324..211S} {324, 211}

\bibitem[\protect\citeauthoryear{Steck}{Steck}{2020}]{Steck2020}
Steck D.~A.,  2020, Quantum and Atom Optics, \url {http://steck.us/teaching}

\bibitem[\protect\citeauthoryear{Szymczak, Olech, Wolak, Bartkiewicz  \&
  Gawroński}{Szymczak et~al.}{2016}]{Szymczak2016}
Szymczak M.,  Olech M.,  Wolak P.,  Bartkiewicz A.,   Gawroński M.,  2016,
  \mn@doi [Monthly Notices of the Royal Astronomical Society: Letters]
  {10.1093/mnrasl/slw044}, 459, L56

\bibitem[\protect\citeauthoryear{{Szymczak}, {Olech}, {Wolak}, {G{\'e}rard}  \&
  {Bartkiewicz}}{{Szymczak} et~al.}{2018}]{Szymczak2018b}
{Szymczak} M.,  {Olech} M.,  {Wolak} P.,  {G{\'e}rard} E.,   {Bartkiewicz} A.,
  2018, \mn@doi [\aap] {10.1051/0004-6361/201833443}, \href
  {http://adsabs.harvard.edu/abs/2018A%26A...617A..80S} {617, A80}

\makeatother
\end{thebibliography}



\appendix

\section{List of Abbreviations}\label{app:abbr}

\begin{description}
    \item[IF:] Integral Fourier
    \item[LMI:] Local Mode Interaction
    \item[MB:] Maxwell-Bloch
    \item[ML:] Menegozzi \& Lamb
    \item[QED:] Quantum Electrodynamics
    \item[SR:] Superradiance
    \item[SVEA:] Slowly-Varying Envelope Approximation
    \item[TML:] Transient Menegozzi \& Lamb

\end{description}

\section{Formal justification for the local mode interaction truncation range}\label{app:lmi_just}

We formally justify the centering of the LMI approximation about $\bar{m} = 0$ on the summation index $\bar{m}$ of equations \eqref{eq:FourierNpm} and \eqref{eq:FourierPpm} via comparison to a perturbative solution in the electric field strength. Consider first a rearrangement of equations \eqref{eq:FourierNpm} and \eqref{eq:FourierPpm} into the form
\begin{align}
    \begin{split}
        \mathbb{N}_{p,m} &= \left(i m d\omega + \frac{1}{T_1}\right)^{-1} \\
        &\quad \times \left[\frac{i}{\hbar} \sum_{\bar{m}} \left(\bar{\mathbb{P}}_{p,\bar{m}}^{+} \mathbb{E}_{p+\bar{m}-m}^{+} - \bar{\mathbb{P}}_{p,\bar{m}}^{-} \mathbb{E}_{p+\bar{m}-m}^{-}\right) + \mathbb{L}_{m}^{(N)}\right] \label{eq:ML_N_Rearr}
    \end{split} \\
    \begin{split}
        \bar{\mathbb{P}}_{p,m}^{+} &= \left(i m d\omega + \frac{1}{T_2}\right)^{-1} \\
        &\quad \times \left[\frac{ i 2 d^{2}}{\hbar} \sum_{\bar{m}} \left(\mathbb{N}_{p,\bar{m}} \mathbb{E}_{p+m-\bar{m}}^{-}\right) + \mathbb{L}_{m}^{(P)}\right]. \label{eq:ML_P_Rearr}
    \end{split}
\end{align}

Let us define vectors of inversion modes and polarisation modes for the $p^\text{th}$ velocity channel as
\begin{equation}
    \vec{N} \equiv
    \begin{pmatrix}
        \vdots \\
        \mathbb{N}_{p,-1} \\ \\
        \mathbb{N}_{p,0} \\ \\
        \mathbb{N}_{p,+1} \\
        \vdots
    \end{pmatrix};
    \quad
    \vec{P} \equiv
    \begin{pmatrix}
        \vdots \\
        \bar{\mathbb{P}}^+_{p,-1} \\ \\
        \bar{\mathbb{P}}^+_{p,0} \\ \\
        \bar{\mathbb{P}}^+_{p,+1} \\
        \vdots
    \end{pmatrix};
    \quad{}
    \vec{x} \equiv
    \begin{pmatrix}
        \vec{N} \\
        \vec{P} \\
        \vec{P}^*
    \end{pmatrix},
\end{equation}
so that equations \eqref{eq:ML_N_Rearr} and \eqref{eq:ML_P_Rearr} may be expressed as
\begin{equation}
    \vec{x} = \overleftrightarrow{M}_E \vec{x} + \vec{b} \label{eq:ML_vec_eqn}
\end{equation}
where
\begin{equation}
    \overleftrightarrow{M}_E = \frac{i}{\hbar}
    \begin{bmatrix}
        \overleftrightarrow{0} & \frac{\overleftrightarrow{E}}{\left(i m d\omega + \frac{1}{T_1}\right)} & -\frac{{\overleftrightarrow{E}}^*}{\left(i m d\omega + \frac{1}{T_1}\right)} \\
        2 d^2 \frac{{\overleftrightarrow{E}}^\dagger}{\left(i m d\omega + \frac{1}{T_2}\right)} & \overleftrightarrow{0} & \overleftrightarrow{0} \\
        -2 d^2 \frac{{\overleftrightarrow{E}}'}{\left(-i m d\omega + \frac{1}{T_2}\right)} & \overleftrightarrow{0} & \overleftrightarrow{0}
    \end{bmatrix}\label{eq:M_E_matrix}
\end{equation}
for submatrices $\overleftrightarrow{E}$ and ${\overleftrightarrow{E}}'$ having elements
\begin{equation}
    \left[\overleftrightarrow{E}\right]_{m,\bar{m}} = \mathbb{E}^+_{p+\bar{m}-m} \text{ and } \left[{\overleftrightarrow{E}}'\right]_{m,\bar{m}} = \left[\overleftrightarrow{E}\right]_{-m,\bar{m}}, \label{eq:E_mat_elts}
\end{equation}
and where $*$ denotes the complex conjugate and $\dagger$ the adjoint. In deriving the submatrix in the third row and first column of equation \eqref{eq:M_E_matrix}, we took the complex conjugate of equation \eqref{eq:ML_P_Rearr}, recognised that $\mathbb{N}^*_{p,\bar{m}}=\mathbb{N}_{p,-\bar{m}}$, and made a change of summation variable $\bar{m} \rightarrow -\bar{m}$. The vector $\vec{b}$ has upper elements
\begin{equation}
    \left[ \vec{b} \right]_{m,\text{upper}} = \frac{\mathbb{L}^{\left(N\right)}_m}{i m d\omega + \frac{1}{T_1}},\label{eq:bupper}
\end{equation}
middle elements
\begin{equation}
    \left[ \vec{b} \right]_{m,\text{middle}} = \frac{\mathbb{L}^{\left(P\right)}_m}{i m d\omega + \frac{1}{T_2}},\label{eq:bmiddle}
\end{equation}
and lower elements
\begin{equation}
    \left[ \vec{b} \right]_{m,\text{lower}} = \frac{{\mathbb{L}^{\left(P\right)}_m}^*}{-i m d\omega + \frac{1}{T_2}}.\label{eq:blower}
\end{equation}

A perturbative solution to equation \eqref{eq:ML_vec_eqn} in increasing powers of the matrix $\overleftrightarrow{M}_E$ (i.e., in increasing powers of the electric field strength) may be identified from inspection to be
\begin{equation}
    \vec{x} = \left(\overleftrightarrow{1} + \overleftrightarrow{M}_E + \overleftrightarrow{M}^2_E + \overleftrightarrow{M}^3_E + \dots \right) \vec{b} \label{eq:pert_soln}
\end{equation}
(to verify, substitute into both sides of equation \eqref{eq:ML_vec_eqn} and observe equality to all orders in $\overleftrightarrow{M}_E$).

We are now in a position to argue that our decision to truncate summations over $\bar{m}$ about the central value $\bar{m}=0$ in Section \ref{subsubsec:lmia} was made in order to maintain consistency with results of first order perturbation in the field strength. Consider equation \eqref{eq:pert_soln} truncated to first order in $\overleftrightarrow{M}_E$. For  pumps constant in the time domain, only their zeroth modes vanish; i.e., $\mathbb{L}^{\left(N / P\right)}_m = \mathbb{L}^{\left(N / P\right)}_0 \delta_{m,0}$. Thus each of expressions \eqref{eq:bupper}--\eqref{eq:blower} contains a non-vanishing value in only its $m=0$ element.

To first order, equation \eqref{eq:pert_soln} acts on each of $\vec{b}_\text{upper}$, $\vec{b}_\text{middle}$, and $\vec{b}_\text{lower}$ with matrix multiplication by first powers of the submatrices $\overleftrightarrow{E}$, ${\overleftrightarrow{E}}^*$, ${\overleftrightarrow{E}}^\dagger$ and ${\overleftrightarrow{E}}'$. Such multiplication involves summation over the $\bar{m}$ column index of expressions \eqref{eq:E_mat_elts}. As per the observation of the prior paragraph, the $\vec{b}$ vectors possess non-vanishing values in only their $m=0$ elements; consequently, matrix multiplication upon them remains correct when the column summation of the matrix multiplication operation is truncated to only the $\bar{m}=0$ column. Comparing expressions \eqref{eq:E_mat_elts} to equations \eqref{eq:ML_N_Rearr} and \eqref{eq:ML_P_Rearr}, we see that this statement concerning matrix multiplication is equivalent to the statement that summations over $\bar{m}$ in equations \eqref{eq:ML_N_Rearr} and \eqref{eq:ML_P_Rearr} be truncated to $\bar{m}=0$.

This constitutes the formal argument that the LMI approximation's summation truncation range should be centered about $\bar{m}=0$ in order to maintain consistency with first order perturbation results; this observation removes the centering ambiguity in the introduction of the LMI approximation in Section \ref{subsubsec:lmia}. Suppose, for example, that we had alternatively expressed the first term on the right side of equation \eqref{eq:FourierNpm} by a less judicious choice of summation variable $m'=\bar{m}-m$, such that
\begin{align}
        \left(i m d\omega \right) \mathbb{N}_{p,m} &= \frac{i}{\hbar} \sum_{\bar{m}} \left(\bar{\mathbb{P}}_{p,\bar{m}}^{+} \mathbb{E}_{p+\bar{m}-m}^{+} - \dots \right.  \\
        &=\frac{i}{\hbar} \sum_{m'} \left(\bar{\mathbb{P}}_{p,m'+m}^{+} \mathbb{E}_{p+m'}^{+} - \dots \right. .
\end{align}
Had we naively proposed that the LMI approximation be achieved by truncating summation about $m'=0$ (i.e., about field modes centered upon the natural Doppler shifted resonance of the velocity channel--a very reasonable proposition), we would have violated consistency with first order perturbation results.

\onecolumn
\section{The integral Fourier representation of the Maxwell-Bloch equations in its real and imaginary parts}\label{app:if_mbes_re_n_im}

We present below the full expression of the IF mode relations in their real and imaginary parts, where all algebra eliminating negative modes of the inversion has been completed. In the summations over $\bar{m}$, the symbol $\mathcal{T}$ denotes the truncated range $\left[-N_\mathrm{int},+N_\mathrm{int}\right]$ and the symbol $\mathcal{T}^+$ denotes the positive truncated range $\left[+1, +N_\mathrm{int}\right]$. Indices $m$ and $p$ span those ranges described in Section \ref{subsubsec:lmia_ifr}.

\begin{align}
    d\omega\left(\mathbb{N}_{p,0}^{\mathbb{R}} - N_{p}^{\mathbb{R}}\left(0\right)\right) &= -\frac{2 d\omega}{\hbar}\sum_{\bar{m}\in\mathcal{T}}\left(\bar{\mathbb{P}}_{p,\bar{m}}^{\mathbb{R}}\Xi_{\bar{m}+p}^{\mathbb{I}} + \bar{\mathbb{P}}_{p,\bar{m}}^{\mathbb{I}}\Xi_{\bar{m}+p}^{\mathbb{R}}\right) -\frac{\pi}{T_{1}}\mathbb{N}_{p,0}^{\mathbb{R}} + \sum_{\bar{m}\in\mathcal{T}^{+}}\frac{2}{\bar{m}T_{1}}\mathbb{N}_{p,\bar{m}}^{\mathbb{I}} + \pi\mathbb{L}_{0}^{(N)\mathbb{R}} - \sum_{m'>0}\mathbb{L}_{m'}^{(N)\mathbb{I}} \\ \nonumber \\
    md\omega\mathbb{N}_{p,m\in\mathcal{T}^{+}}^{\mathbb{R}} &= \frac{1}{\hbar}\sum_{\bar{m}\in\mathcal{T}}\left[\bar{\mathbb{P}}_{p,\bar{m}}^{\mathbb{R}}\left(\mathbb{E}_{\bar{m}-m+p}^{\mathbb{R}} - \mathbb{E}_{\bar{m}+m+p}^{\mathbb{R}}\right) + \bar{\mathbb{P}}_{p,\bar{m}}^{\mathbb{I}}\left(\mathbb{E}_{\bar{m}+m+p}^{\mathbb{I}} - \mathbb{E}_{\bar{m}-m+p}^{\mathbb{I}}\right)\right] -  \frac{\mathbb{N}_{p,m}^{\mathbb{I}}}{T_{1}} + \mathbb{L}_{m}^{(N)\mathbb{I}} \\ \nonumber \\
    \begin{split}
    md\omega\mathbb{N}_{p,m\in\mathcal{T}^{+}}^{\mathbb{I}} &= \frac{1}{\hbar}\sum_{\bar{m}\in\mathcal{T}}\left[\bar{\mathbb{P}}_{p,\bar{m}}^{\mathbb{R}}\left(\mathbb{E}_{\bar{m}+m+p}^{\mathbb{I}} + \mathbb{E}_{\bar{m}-m+p}^{\mathbb{I}} - 2\mathbb{E}_{\bar{m}+p}^{\mathbb{I}}\right) + \bar{\mathbb{P}}_{p,\bar{m}}^{\mathbb{I}}\left(\mathbb{E}_{\bar{m}+m+p}^{\mathbb{R}} + \mathbb{E}_{\bar{m}-m+p}^{\mathbb{R}} - 2\mathbb{E}_{\bar{m}+p}^{\mathbb{R}}\right)\right] \\
    &\qquad + \frac{1}{T_{1}}\left(\mathbb{N}_{p,m}^{\mathbb{R}} - \mathbb{N}_{p,0}^{\mathbb{R}}\right) + \mathbb{L}_{0}^{(N)\mathbb{R}} - \mathbb{L}_{m}^{(N)\mathbb{R}}
    \end{split} \\ \nonumber \\
    \begin{split}
    d\omega\left(\bar{\mathbb{P}}_{p,0}^{\mathbb{R}} - \bar{\mathcal{P}}_{p}^{\mathbb{R}}\left(0\right)\right) &= \frac{2d^{2} d\omega}{\hbar}\left\{\mathbb{N}_{p,0}^{\mathbb{R}}\Xi_{p}^{\mathbb{I}} + \sum_{\bar{m}\in\mathcal{T}^{+}}\left[\mathbb{N}_{p,\bar{m}}^{\mathbb{R}}\left(\Xi_{p-\bar{m}}^{\mathbb{I}} + \Xi_{p+\bar{m}}^{\mathbb{I}}\right) + \mathbb{N}_{p,\bar{m}}^{\mathbb{I}}\left(\Xi_{p+\bar{m}}^{\mathbb{R}}-\Xi_{p-\bar{m}}^{\mathbb{R}}\right)\right]\right\} \\
    &\qquad - \frac{\pi}{T_{2}}\bar{\mathbb{P}}_{p,0}^{\mathbb{R}} + \sum_{\bar{m}\in\mathcal{T}^{\pm}}\frac{1}{\bar{m}T_{2}}\bar{\mathbb{P}}_{p,\bar{m}}^{\mathbb{I}} + \pi\mathbb{L}_{0}^{(P)\mathbb{R}} - \sum_{m'\neq0}\frac{1}{m'}\mathbb{L}_{m'}^{(P)\mathbb{I}}
    \end{split} \\ \nonumber \\
    \begin{split}
    d\omega\left(\bar{\mathbb{P}}_{p,0}^{\mathbb{I}} - \bar{\mathcal{P}}_{p}^{\mathbb{I}}\left(0\right)\right) &= \frac{2d^{2} d\omega}{\hbar}\left\{\mathbb{N}_{p,0}^{\mathbb{R}}\Xi_{p}^{\mathbb{R}} + \sum_{\bar{m}\in\mathcal{T}^{+}}\left[\mathbb{N}_{p,\bar{m}}^{\mathbb{R}}\left(\Xi_{p-\bar{m}}^{\mathbb{R}} + \Xi_{p+\bar{m}}^{\mathbb{R}}\right) + \mathbb{N}_{p,\bar{m}}^{\mathbb{I}}\left(\Xi_{p-\bar{m}}^{\mathbb{I}} - \Xi_{p+\bar{m}}^{\mathbb{I}}\right)\right]\right\} \\
    &\qquad - \frac{\pi}{T_{2}}\bar{\mathbb{P}}_{p,0}^{\mathbb{I}} + \sum_{\bar{m}\in\mathcal{T}^{\pm}}\frac{1}{\bar{m}T_{2}}\bar{\mathbb{P}}_{p,\bar{m}}^{\mathbb{R}} + \pi\mathbb{L}_{0}^{(P)\mathbb{I}} + \sum_{m'\neq0}\frac{1}{m'}\mathbb{L}_{m'}^{(P)\mathbb{R}} 
    \end{split} \\ \nonumber \\
    \begin{split}
    md\omega\bar{\mathbb{P}}_{p,m}^{\mathbb{R}} &= \frac{2d^{2}}{\hbar}\left\{\mathbb{N}_{p,0}^{\mathbb{R}}\left(\mathbb{E}_{m+p}^{\mathbb{R}} - \mathbb{E}_{p}^{\mathbb{R}}\right) + \sum_{\bar{m}\in\mathcal{T}^{+}}\left[\mathbb{N}_{p,\bar{m}}^{\mathbb{R}}\left(\mathbb{E}_{m+\bar{m}+p}^{\mathbb{R}} + \mathbb{E}_{m-\bar{m}+p}^{\mathbb{R}} - \mathbb{E}_{\bar{m}+p}^{\mathbb{R}} - \mathbb{E}_{-\bar{m}+p}^{\mathbb{R}}\right)\right.\right. \\
    &\qquad + \left.\left. \mathbb{N}_{p,\bar{m}}^{\mathbb{I}}\left(\mathbb{E}_{m-\bar{m}+p}^{\mathbb{I}} + \mathbb{E}_{\bar{m}+p}^{\mathbb{I}} - \mathbb{E}_{m+\bar{m}+p}^{\mathbb{I}} - \mathbb{E}_{-\bar{m}+p}^{\mathbb{I}}\right)\right]\vphantom{\sum_{\bar{m}\in\mathcal{T}^{+}}}\right\} + \frac{1}{T_{2}}\left(\bar{\mathbb{P}}_{p,0}^{\mathbb{I}} - \bar{\mathbb{P}}_{p,m}^{\mathbb{I}}\right) + \mathbb{L}_{m}^{(P)\mathbb{I}} - \mathbb{L}_{0}^{(P)\mathbb{I}}
    \end{split} \\ \nonumber \\
    \begin{split}
    md\omega\bar{\mathbb{P}}_{p,m}^{\mathbb{I}} &= \frac{2d^{2}}{\hbar}\left\{\mathbb{N}_{p,0}^{\mathbb{R}}\left(\mathbb{E}_{p}^{\mathbb{I}} - \mathbb{E}_{m+p}^{\mathbb{I}}\right) + \sum_{\bar{m}\in\mathcal{T}^{+}}\left[\mathbb{N}_{p,\bar{m}}^{\mathbb{R}}\left(\mathbb{E}_{\bar{m}+p}^{\mathbb{I}} + \mathbb{E}_{-\bar{m}+p}^{\mathbb{I}} - \mathbb{E}_{m+\bar{m}+p}^{\mathbb{I}} - \mathbb{E}_{m-\bar{m}+p}^{\mathbb{I}}\right)\right.\right. \\
    &\qquad + \left.\left.\mathbb{N}_{p,\bar{m}}^{\mathbb{I}}\left(\mathbb{E}_{m-\bar{m}+p}^{\mathbb{R}} + \mathbb{E}_{\bar{m}+p}^{\mathbb{R}} - \mathbb{E}_{m+\bar{m}+p}^{\mathbb{R}} - \mathbb{E}_{-\bar{m}+p}^{\mathbb{R}}\right)\right]\vphantom{\sum_{\bar{m}\in\mathcal{T}^{+}}}\right\} + \frac{1}{T_{2}}\left(\bar{\mathbb{P}}_{p,m}^{\mathbb{R}} - \bar{\mathbb{P}}_{p,0}^{\mathbb{R}}\right) + \mathbb{L}_{0}^{(P)\mathbb{R}}-\mathbb{L}_{m}^{(P)\mathbb{R}}.
    \end{split}
\end{align}


\bsp	
\label{lastpage}
\end{document}